\documentclass[12pt,preprint]{aastex}

\usepackage{emulateapj5}

\slugcomment{Accepted for publication in ApJ}

\shorttitle{Multiwavelength Star Formation Rates}
\shortauthors{Cardiel et al.}

\begin{document}


\title{A Multiwavelength Approach to the SFR Estimation \\
in Galaxies at intermediate redshifts$^{\rm 1,2,3}$}


\author{N. Cardiel\altaffilmark{4},
        D. Elbaz\altaffilmark{5,6},
        R. P. Schiavon,
        C. N. A. Willmer\altaffilmark{7},
        D. C. Koo,
        A. C. Phillips, and
        J. Gallego\altaffilmark{4}}
\email{ncl@astrax.fis.ucm.es}
\email{elbaz@cea.fr}
\email{ripisc, cnaw, koo, phillips@ucolick.org}
\email{jgm@astrax.fis.ucm.es}
\affil{UCO/Lick Observatory, University of California, 
Santa Cruz, CA 95064}

\footnotetext[1]{Based on data obtained at the W. M. Keck Observatory, which is
operated as a scientific partnership among the California Institute of
Technology, the University of California, and the National Aeronautics and
Space Administration. The Observatory was made possible by the generous
financial support of the W. M. Keck Foundation.}
\footnotetext[2]{Based in part on observations with the NASA/ESA {\sl Hubble
Space Telescope\/}, obtained at the Space Telescope Science Institute, which is
operated by the Association of Universities for Research in Astronomy, Inc., 
under NASA contract NAS 5-26555.}
\footnotetext[3]{Based in part on observations with ISO, an ESA project with
instruments funded by ESA Member States (especially the PI countries: France,
Germany, the Netherlands and the United Kingdom) with the participation of ISAS
and NASA.}
\altaffiltext{4}{Departamento de Astrof\'{\i}sica, Facultad de F\'{\i}sicas,
Avenida Complutense s/n, 28040-Madrid, Spain}
\altaffiltext{5}{CEA Saclay - Service d'Astrophysique, Orme des Merisiers,
91191 Gif-sur-Yvette C\'{e}dex, France}
\altaffiltext{6}{Physics Department, University of California, Santa Cruz, CA
95064}
\altaffiltext{7}{On leave from Observat\'{o}rio Nacional, Rua General 
Jos\'{e} Cristino 77, S\~{a}o Cristov\~{a}o, Rio de Janeiro, Brazil}

\begin{abstract}
We use a sample of 7 starburst galaxies at intermediate redshifts ($z \sim 0.4$
and $z \sim 0.8$) with observations ranging from the observed ultraviolet to
1.4 GHz, to compare the star formation rate (SFR) estimators which are used in
the different wavelength regimes. We find that {\it extinction corrected\/}
H$\alpha$ underestimates the SFR, and the degree of this underestimation
increases with the infrared luminosity of the galaxies.  Galaxies with very
different levels of dust extinction as measured with SFR$_{\rm IR}$/SFR(${\rm H
\alpha}$,uncorrected for extinction) present a similar attenuation
A[H$\alpha$], as if the Balmer lines probed a different region of the galaxy
than the one responsible for the bulk of the IR luminosity for large SFRs.  In
addition, SFR estimates derived from [O~{\sc ii}]$\lambda$3727 match very well
those inferred from H$\alpha$ after applying the metallicity correction derived
from local galaxies. SFRs estimated from the UV luminosities show a
dichotomic behavior, similar to that previously reported by other authors in
galaxies at $z \la 0.4$. Here we extend this result up to $z \sim 0.8$.
Finally, one of the studied objects is a luminous compact galaxy (LCG) that may
be suffering similar dust-enshrouded star formation episodes.  These results
highlight the relevance of quantifying the actual $L_{\rm IR}$ of LCGs, as well
as that of a much larger and generic sample of luminous infrared galaxies,
which will be possible after the launch of SIRTF.
\end{abstract}


\section{INTRODUCTION}
\label{introduction}

The observational efforts devoted during the last years to measure the
cosmic star formation history is clearly manifest in the growing
number of works in the field.  As a result of the advent of the Hubble
Space Telescope and of the 8-10 meter class telescopes, the rest-frame
ultraviolet (UV) star formation history per unit comoving volume
initially derived up to $z~\sim$ 1 \citep{lil96} was extended up to
$z\,\sim$ 4 \citep{mad96,ste99}.  Naturally, all these works rely on
the use of the available star formation rate (SFR) diagnostics:
nebular emission lines like H$\alpha$
\citep{ken92a,ken98,gal95,tre98,gla99,moo00,hop00} and [O~{\sc ii}]
\citep{hog98,ara02}; UV continuum luminosities
\citep{lil96,mad96,try98,cow99,ste99,sul00,sul01}; far-infrared
luminosities \citep{row97,bla99,flo99,cha02}; and 1.4~GHz radio
luminosities \citep{con92,cra98}. Different cosmic star formation
histories have been advocated from these various studies which can
only be reconciled if the effect of dust extinction is accounted for.

In the local universe, galaxies radiate the bulk of their luminosity in the
stellar regime, i.e. below $\lambda \sim$ 5\,$\mu$m. On average, only about one
third of their stellar light is absorbed by dust and re-emitted in the dust
regime from the mid infrared (MIR) to the sub-millimeter \citep{soi91}.  In the
distant universe, the reverse seems to take place as suggested by the combined
results of the COsmic Background Experiment (COBE), the ISOCAM and ISOPHOT
instruments on board the Infrared Space Observatory (ISO) and the
sub-millimeter camera SCUBA at the JCMT. Number counts at 15\,$\mu$m
\citep{elb99}, at 90\,$\mu$m and 170\,$\mu$m \citep{efs00,kaw00,mat00,dol01},
and at 450 and 850\,$\mu$m \citep[][and references therein]{sma02} revealed the
presence of an excess of faint galaxies by one order of magnitude in comparison
with expectations based on the local properties of galaxies. Such excess can
only be explained if galaxies emitted a larger fraction of their luminosity in
the dust regime in the past. This conclusion is reinforced by the detection of
a strong cosmic infrared background (CIRB) above $\lambda\sim$ 100\,$\mu$m
\citep{pug96,hau98,fix98,lag99,lag00} the bulk of which appears to be produced
by starbursting galaxies radiating the majority of their light in the IR above
$\lambda \sim$ 5\,$\mu$m \citep{elb02,cha02}.

Two major questions remain unanswered:

-- How consistent are the different SFR indicators commonly used
in the literature? 

-- Do local and distant galaxies exhibit a similar behaviour?

Studies of local galaxies have shown that even after correcting for
dust extinction, a significant discrepancy remains between SFRs
derived from UV and H$\alpha$ measurements compared with those
obtained from FIR and radio luminosities, and that this discrepancy is
worse for galaxies with higher SFRs \citep{hop01,pog00}. \citet{hop01}
and \citet{sul01} report an increase of the
$F$(H$\alpha$)/$F$(H$\beta$) ratio, used to measure extinction, with
the SFR of galaxies. They suggest to use the fit to this correlation
in order to account for dust extinction when H$\beta$ is not
available. However, \citet{bua02} indicate that the apparent link
between extinction and star formation is likely not real, but only the
result of a dispersed correlation between dust extinction and
luminosity.

In a recent paper, \citet{rig00} studied a sample of 12 galaxies with
$0.4 < z < 1.4$, detected with ISOCAM in the Hubble Deep Field South
(HDF-S). They showed that the discrepancy found between optical-UV and
FIR-radio SFRs in the local universe is also present at high~$z$.
However, the Balmer decrement was not available for these galaxies and
the authors discussed the effect of dust extinction assuming an
average correction factor of 4 for SFR(H$\alpha$).

In the present paper, we propose to address the two previous questions
using a test sample of seven galaxies at $z\sim$ 0.4 and 0.8, five of
which have MIR flux densities from which we derived FIR luminosities
assuming that the correlation between MIR and FIR luminosities
observed in the local universe remains valid at these redshifts
\citep{elb02,cha02}. The galaxies were selected from their MIR flux
density in order to sample SFRs in the range from 2 to
160\,M$_{\odot}$ yr$^{-1}$ and morphologies ranging from a normal
spiral to merging galaxies.

In this work we extend to higher redshifts the comparison between
different SFR estimators that have been used for samples of local
galaxies (e.g., Sullivan et al. 2001). For that purpose, we compare
classical SFR indicators such as the nebular emission lines [O~{\sc
ii}] and H$\alpha$, or the UV continuum, that we correct for dust
extinction using the Balmer decrement, with the FIR and radio
luminosities which are unaffected by extinction. This has been made
possible with the comissioning of high resolution optical and
near-infrared (NIR) spectroscopy on the Keck telescopes. High
resolution spectroscopy is required not only to avoid the
contamination from atmospheric OH emission lines but also to resolve
the [N~{\sc ii}]$\lambda\lambda6549,6583$~\AA\ and H$\alpha$ emission
lines. After the presentation of the observational data in
Sect.~\ref{data}, the fluxes in these three lines will be combined in
Sect.~\ref{agn} with the [O~{\sc iii}] and H$\beta$ lines to locate
galaxies in diagnostic diagrams that allow separating between galaxies
whose optical luminosity is mainly due to star formation, from
galaxies where an Active Galactic Nucleus is the dominant source of
emission.

In Sect.~\ref{sfr} we obtain multiwavelength SFR estimates for the
galaxy sample using the available flux data. Additional galaxy
parameters are examined in Sect.~\ref{galaxy_properties}, and a
summary of the main properties of each object is presented in
Sect.~\ref{description_individual_galaxies}. In
Sect.~\ref{sfr_comparison} we compare and discuss the results obtained
using the different SFR estimators. Finally, we present the conclusions in
Sect.~\ref{conclusions}.

Unless otherwise indicated, throughout this paper we will assume $H_0 = 70 \;
{\rm km} \; {\rm s}^{-1} \; {\rm Mpc}^{-1}$, $\Omega_M=0.3$, and
$\Omega_\Lambda=0.7$.

\section{THE DATA}
\label{data}
\subsection{Galaxy Sample and Observations}

\begin{figure}
\epsscale{0.70}
\plotone{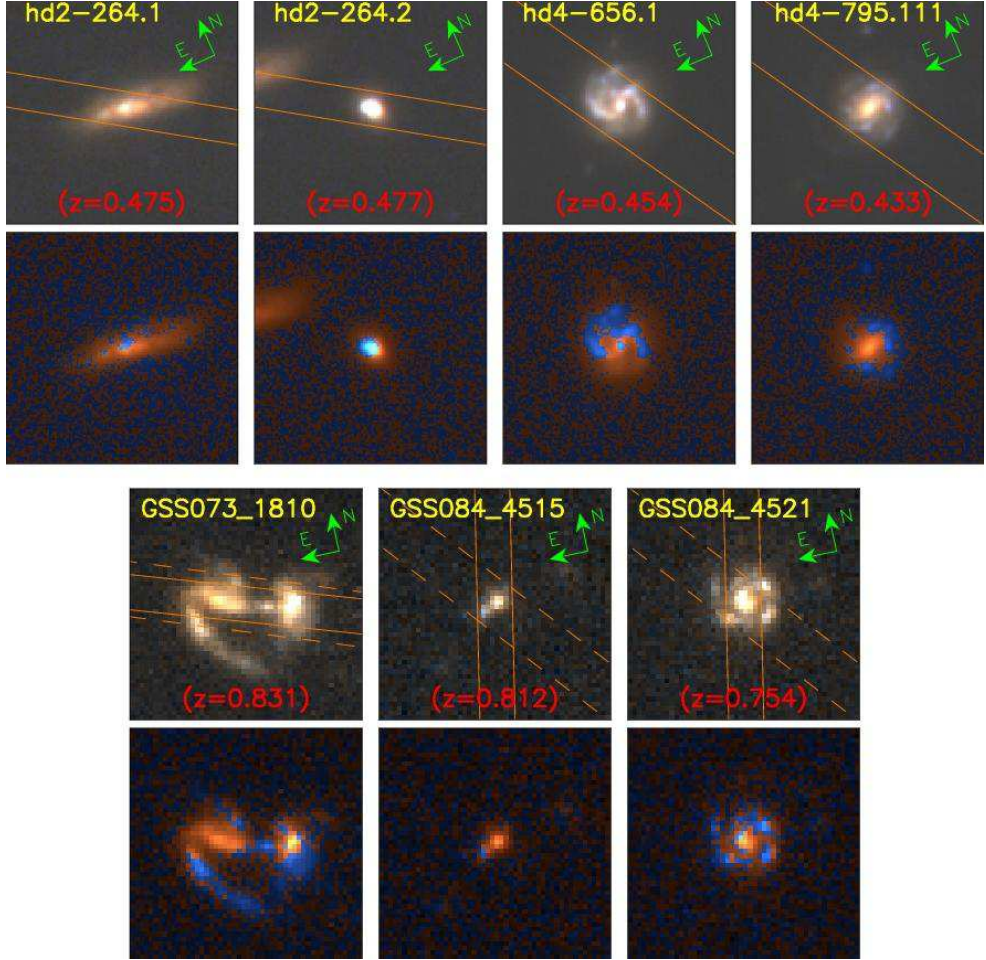}
\figcaption{HST images of the whole galaxy sample. Each panel corresponds to
$5\arcsec \times 5 \arcsec$. The upper panel of each object is a composite
$RGB$ image derived from the combination of the flux calibrated images.  For
the HDF-N galaxies, the average of F300W and F450W was employed as $B$, F606W
as $G$, and F814W as $R$. For the GSS galaxies only F606W and F814W were
available. For that reason, the former was employed as $B$, the latter as $R$,
and the average of both as $G$. The lower panels are blue-and-red enhanced
compositions, in which the reddest and bluest regions have been intentionally 
exaggerated. For that purpose, all the images for each galaxy were scaled to
the same total averaged flux {\sl per unit wavelength\/} prior to the color
composition (in this sense, differential color variations are easier to
display). Next, images were assigned to $RGB$ using the same criterion than
with the upper panels, but before combining them, the weighting factors
$(1,1/2,1/4)$ and $(1/4,1/2,1)$ were applied to the $(R,G,B)$ set of each pixel
which held $R>B$ and $R<B$, respectively. For the HDF-N galaxies the solid
lines indicate the orientation and width of the slit in the observation with
ESI. For the GSS galaxies the solid and dashed lines correspond to the slit
employed in the observation with NIRSPEC and LRIS, respectively. We remind the
readers that total exposure times (in seconds) for HDF-N images are  153700
(F300W), 120600 (F450W), 109050 (F606W), and 123600 (F814W).  For GSS073 these
numbers are 24400 (F606W) and 25200 (F814W), whereas the integrations in GSS084
are the shortest, 2800 (F606W), and 4400 (F814W).  \label{figure_mosaic}}
\end{figure}

The sample of dusty starburst galaxies was chosen according to the availability
of ISOCAM-15~$\mu$m luminosities (or, at least, upper limits) for each object.
The galaxies were also constrained to have a spectroscopic redshift accurate
enough to insure that the H$\alpha$ line would be free of contamination due to
night sky lines. With those criteria in mind, we have selected 4~galaxies with
a mean redshift $\langle z \rangle \sim 0.46$ in the Hubble Deep Field North,
\mbox{HDF-N} \citep{wil96}, and 3~galaxies with $\langle z \rangle \sim 0.80$
in the Groth Strip Survey, GSS \citep{gro94,koo96}. The whole set of galaxies
is listed in Table~\ref{table_sample} and displayed in
Fig.~\ref{figure_mosaic}. Interestingly, the two galaxies undetected by ISO
have a compact morphology (hd2-264.2 and GSS084\_4515). The ISO sources are
face-on spirals (hd4-656.1, hd4-795.111, and GSS084\_4521), a disky galaxy
(hd2-264.1), and a colliding system (GSS073\_1810) which looks like a higher
redshift analogue of the local Antenae system (NGC~4038 and NGC~4039). In order
to facilitate the recognition of the individual objects in some of the figures
of this paper, we have labeled the symbols with the letters S, D, C and A,
corresponding to their morphologies, i.e.\ Spiral, Disky, Compact and
Antenae-like, respectively.

According to their IR luminosities (computed as described in
Sect.~\ref{subsection_sfr_ir} and listed in
Table~\ref{table_more_galaxy_properties}), the HDF-N galaxies are not luminous
enough to be classified as luminous infrared galaxies (LIRG, $L_{\rm IR} \ge
10^{11} L_\odot$), whereas the contrary is true for the GSS objects. Note that
although GSS073\_1810 could be marginally considered as an ultraluminous
infrared galaxy (ULIRG; $L_{\rm IR} \ge 10^{12} L_\odot$), this object is
constituted by two colliding galaxies (see Fig.~\ref{figure_mosaic}). A more
detailed characterization of the galaxy sample is presented in
Sects.~\ref{galaxy_properties} and~\ref{description_individual_galaxies}.

For all galaxies in our sample, we have obtained optical and near-infrared
spectroscopy. The observations were carried out with the W.~M.~Keck Observatory
telescopes and 3 different instruments over 5 observing runs (see
Table~\ref{table_sample}). For the relatively low redshift \mbox{HDF-N}
galaxies, we employed the Echelle Spectrograph and Imager, ESI
\citep{sut97,epp98,she02}, which covers the full wavelength range from
3900~\AA\ to 11000~\AA\ in one configuration, and with a spectral resolution
${\rm FWHM} \sim 0.8$~\AA\ at observed $\lambda({\rm H}\alpha) \sim 9600$~\AA.
However, for the more distant GSS galaxies, initially only observed in the
spectral range 4500--9000~\AA\ with the Low-Resolution Imaging Spectrometer,
LRIS \citep{oke95} (${\rm FWHM}\sim 3$~\AA), we expanded the spectral window by
using the near-infrared echelle spectrograph NIRSPEC \citep{mcl98} in the J
band (N2 and N3 filters, total spectral range 1.089-1.375~$\mu{\rm m}$, ${\rm
FWHM} \sim 8$~\AA). In general the observations were carried out with airmasses
ranging from 1.09 to 1.56.

\subsection{Data Reduction and Error Handling}

The reduction of the LRIS data (runs 1, 2 and~3) is as described in
\citet{sim99}. The absolute flux calibration used observations of the
spectrophotometric standard \mbox{BD+33 2642} (runs~1 and~3) as the primary
calibrator and the spectra of two stars, GSS063\_0445 and GSS093\_2453 (runs 1,
2 and~3), used to align the image masks, as secondary calibrators. We find that
the response curve of LRIS is very stable on a run-to-run basis, with small
fluctuations $\lesssim 5$\%, which translate into negligible variations in the
star formation rates once other sources of error and uncertainties are
considered.

The ESI (run~4) and NIRSPEC (run~5) spectra have been reduced using a suite of
our own programs based on the {\sc reduceme} package \citep{car99}, and a new
FORTRAN package, {\sc xnirspec}, specially developed for the reduction of the
near-IR data. Error frames have been computed directly from the raw images,
using the read-out noise and the gain of each detector. The reduction in
parallel of these error frames together with the data images guarantees the
correct propagation of errors due to arithmetic manipulation throughout the
reduction procedure \citep{car98,car02}. In this sense, each fully processed
spectrum is accompanied by an associated error spectrum which contains the
random error in each pixel due to photon statistics and read-out noise.

One of the major problems encountered when measuring spectral features in the
near-IR is the proper removal of the OH emission lines \citep{rou00} and the
compensation for telluric absorption due to water vapour and other molecules
\citep{ste94,chm00}. Systematic variations in the spectral direction (e.g.\
sampling aliasing, inacurate geometric distortion corrections), and in flux
calibration (e.g.\ even small flatfield biases, residual fringing, or slit
width inhomogeneities), pose a difficult challenge to the data reduction.
Trying to minimize part of these problems, the NIRSPEC observations were
carried out nodding the telescope, changing the position of the galaxies on the
slit, in order to feed the data reduction pipeline with the subtraction of
consecutive images. Even when following this procedure, the unpredictable
residuals of each sky line (either in intensity or sign) had to be subtracted
like normal sky lines in a subsequent step.

Since the spectra obtained in the detector plane of ESI and NIRSPEC are
distorted, the image rectification prior to the removal of sky lines generates
aliasing artifacts (the undersampled sky line residuals cannot be properly
rectified), and, in addition, introduces a correlation between adjacent pixels
in the error frames. In order to alleviate both effects, we mapped the image
distortions in the spatial and spectral directions by using bivariate
polynomial transformations \citep{wol90}.  These mapping functions were
determined by fitting the spectrum of a bright star, observed at different
positions along the slit, and to the sky lines.

For the ESI observations, the removal of the telluric absorption, as well as
the absolute flux calibration, used the high signal-to-noise ratio (S/N per \AA\
$\gtrsim 150$) almost featureless spectra of the spectrophotometric stars
(\mbox{BD+28 4211}, Wolf~1346 and Feige~34).  In the NIRSPEC run only one
spectrophotometric star was observed (\mbox{BD+28 4211}), and the removal of
telluric features was performed using the observation of V986~Oph, a hot
(B0III) and bright (${\rm V} = 6.15$~mag) star.

\begin{figure}
\epsscale{1.00}
\plotone{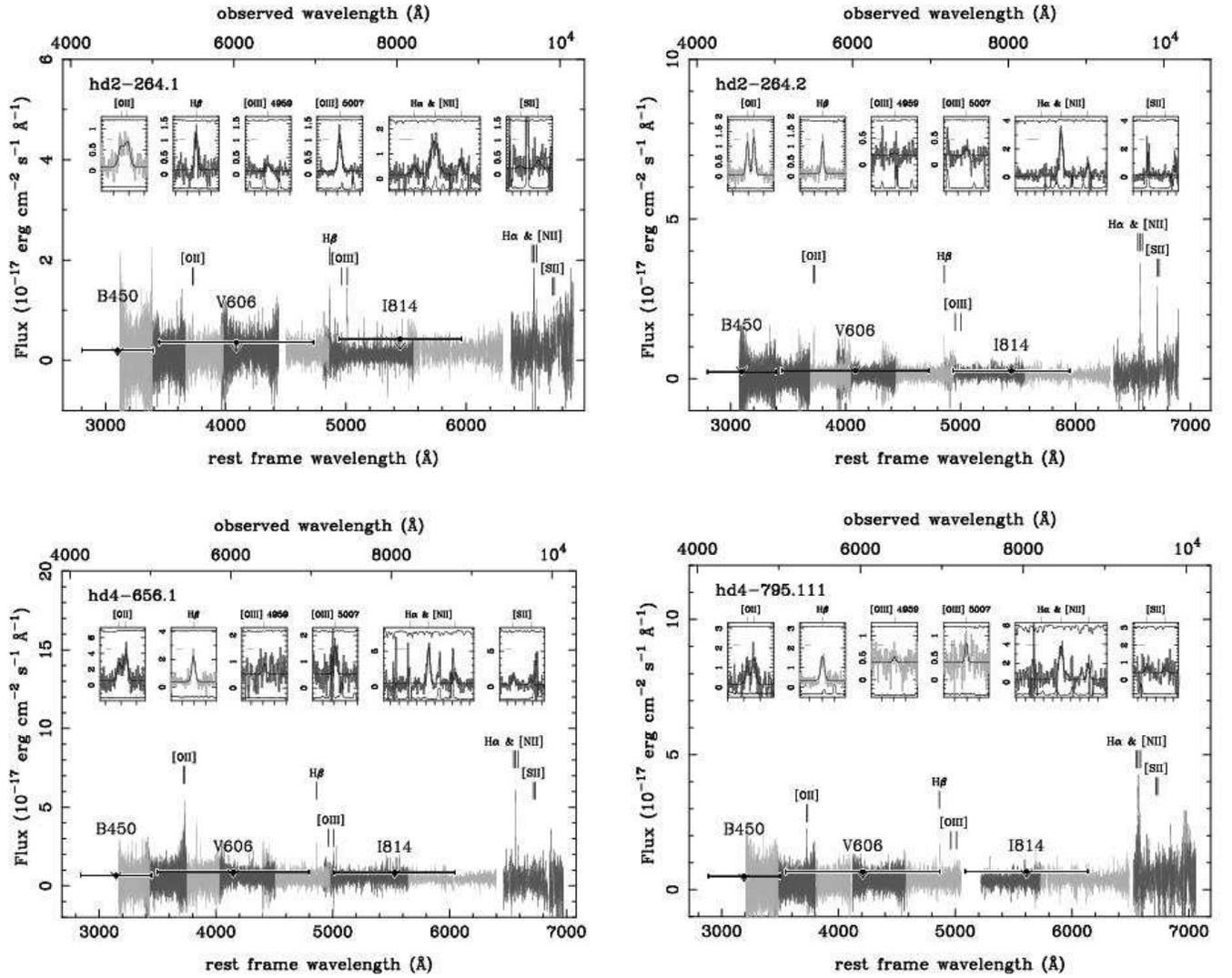}
\figcaption{Flux calibrated spectra for the HDF-N galaxy subsample. Each gray
scale intensity identifies a different order in the ESI echelle spectra.
Very high sky
residuals have been masked only for display purposes. The location of the
detected emission lines is marked. The insets show the fits to each emission
line (in these enlargements sky residuals have not been masked).
Inside these panels we have also plotted the normalized atmospheric
transmission (upper curve, with the zero level marked by the short horizontal
line in the upper left side), the subtracted sky spectrum (lower curve, in
arbitrary units), and the resulting emission line fit (thin line overplotted on
spectrum). The zoomed wavelength range is 1600 km/s (3200 km/s for the
H$\alpha$ \& [N~{\sc ii}] panel), with ticks every 10~\AA\ (observed
wavelength). The broad-band photometry collected in
Table~\ref{table_more_fluxes_hdfn} is also displayed. Error bars in the
spectral direction represent the filter coverage, whereas the arrows indicate
the decrease in flux after applying the aperture corrections given in
Table~\ref{table_aperture_corrections}. There is a good agreement in the
absolute flux calibration of the spectroscopic and broad-band photometry data.
The residual fringing present in the ESI data is responsible for the high
residuals in the H$\alpha$ and [S~{\sc ii}] regions.\label{figure_sp_hdfn}}
\end{figure}

\begin{figure}
\epsscale{1.00}
\plotone{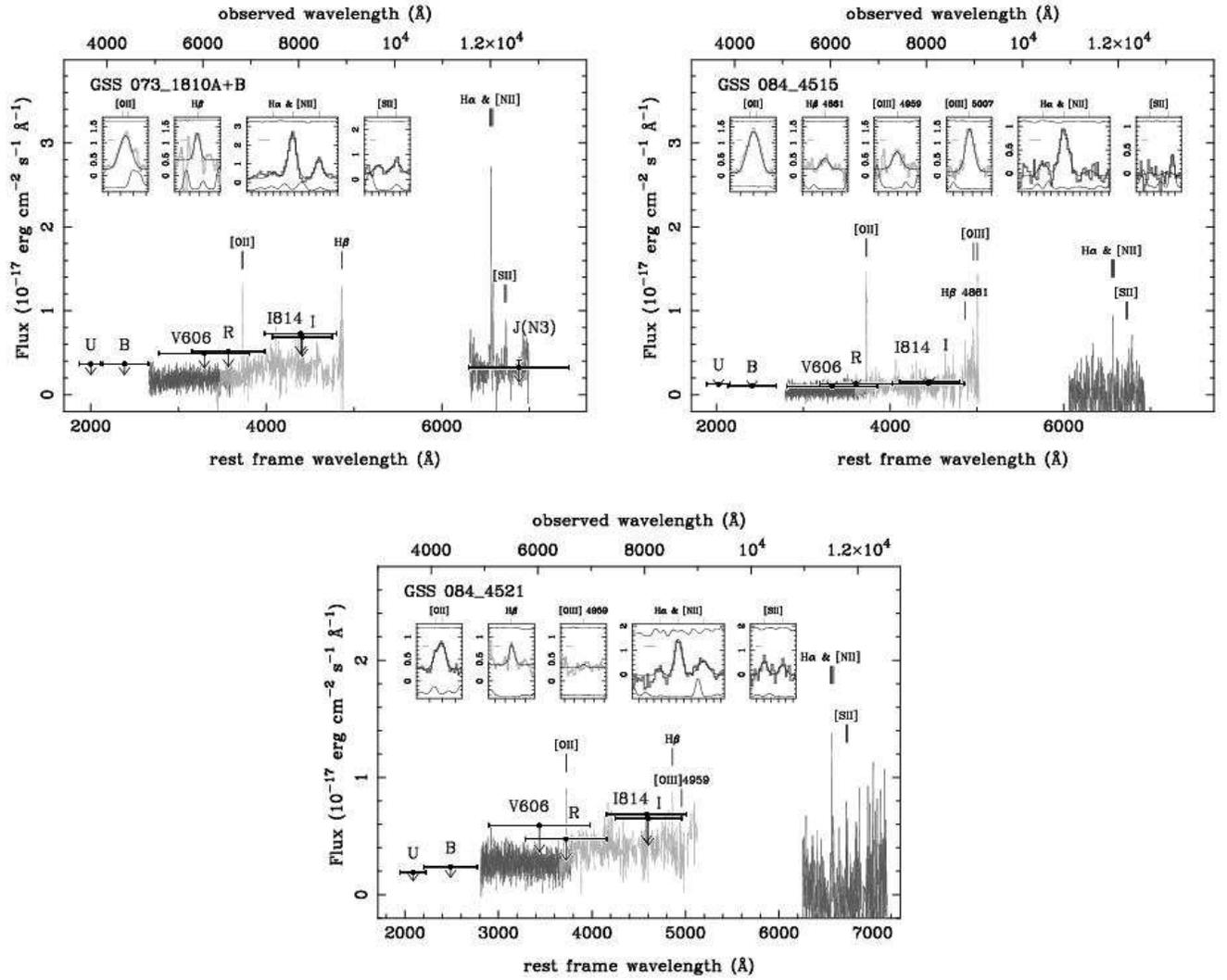}
\figcaption{Flux calibrated spectra for the GSS galaxy subsample. 
In this case, LRIS spectra are plotted together with NIRSPEC data. See
caption of Fig.~\ref{figure_sp_hdfn} for explanation. The broad-band photometry
corresponds to data collected in Table~\ref{table_more_fluxes_gss}. Note that
for these objects, the agreement in the absolute flux calibration of the
spectroscopic and broad-band photometry data is also good, exception for the
$J_{\rm N3}$ value of GSS073\_1810. However, in this case the uncertainty is
much larger than in the rest of the broad-band measurements (note that we have
plotted an upper error bar segment in flux for this point).
\label{figure_sp_gss}}
\end{figure}

All the flux calibrated spectra are displayed in Figs.~\ref{figure_sp_hdfn}
and~\ref{figure_sp_gss}.

\subsection{Measure of emission line fluxes}

The spectra have been corrected for atmospheric extinction using the Mauna Kea
values available on the WWW page of the UKIRT (United Kingdom Infrared
Telescope). The small corrections for galactic extinction used the dust maps of
\citet{sch98} and the galactic extinction curve of \citet{fit99}.

With the aim of correcting the measured Balmer emission line fluxes for the
expected underlying stellar absorption, we have applied the criterion followed
by \citet{cal94}, who, based on the extensive analysis of giant extragalactic
HII regions by \citet{mcc85}, employed ${\rm EW}_{\rm abs}({\rm H}\alpha) =
{\rm EW}_{\rm abs}({\rm H}\beta) = 2$~\AA\ for galaxies with undetected or
uncertain H$\gamma$ flux.

In order to follow a homogeneous procedure in the determination of line
luminosities, line fluxes have been derived from Gaussian fits to the
corresponding emission lines. It is worth noting that these fits not only match
the observed spectra within the error bars available in each pixel, but they
also help minimizing the uncertainties due to the choice of integration limits
on both sides of the emission lines.  Flux errors are also easily computed in
this way through numerical simulations with bootstrapped error spectra.  All
the measured emission lines are displayed in the insets of
Figs.~\ref{figure_sp_hdfn} and~\ref{figure_sp_gss} (see caption for details).
Their values are listed in Table~\ref{table_fluxes}.  Considering that the
H$\alpha$ line has been observed in spectral regions densely populated by sky
lines, it is not surprising that in some cases large residuals prevented
fitting Gaussians to the entire wavelength domain of the [N~{\sc ii}]
$\lambda\lambda$6549, 6583 lines.  In these cases (and also when no clear
signature of these lines was detectable), since the expected location of the
lines is very well constrained by the centroid of the H$\alpha$ line,
restricted fits were performed to the line regions free of contamination (see
caption of Table~\ref{table_fluxes} for details). Restricted fits have also
been applied to other emission lines when required.

Intrinsic reddenings, parameterized through the color excess in the nebular
gas, E(B$-$V)$_{\rm gas}$, were determined using the measured Balmer decrement,
and assuming typical gas conditions, i.e.\ $N_{\rm e}=10^2$~cm$^{-3}$, $T_{\rm
e}=10^4$~K, recombination case~B, and the flux ratio H$\alpha$/H$\beta$=2.86
\citep{ost89}.  Since the extinction properties of the Milky Way
\citep{sav79,sea79,car89,odo94,fit99}, the Small Magellanic Cloud
\citep{pre84,bou85}, and the Large Magellanic Cloud \citep{how83,fit85} are
similar in the optical and in the near-IR spectral ranges, it is common to
assume that the selective extinction curve from any of these objects can be
safely used to correct the emission lines for internal reddening \citep{cal97}.
For this reason, here we adopt the average curve for Galactic extinction
published by \citet{fit99}. The E(B$-$V)$_{\rm gas}$ values for our galaxy
sample are given in the last column of Table~\ref{table_fluxes}. The quoted
errors correspond to the propagation of random errors in the H$\alpha$ and
H$\beta$ fluxes.

Finally, to estimate the impact of the slit widths and position angles in the
measured emission line fluxes, we computed the fraction of light passing
through the slit for each observation, using the HST images; these fractions
and associated errors, are listed in Table~\ref{table_aperture_corrections}.
The errors were calculated using numerical simulations which assume random
positioning errors when placing the slits over the targets.  The simulations
were performed after isolating each galaxy from its neighbors using SExtractor
\citep{ber96}, with the configuration parameter {\sc checkimage\_type} set to
{\sc objects}. Although those slit-loss estimates must be taken with caution,
it seems unlikely that we are losing more than 50\% of the light in any case.

\subsection{Additional Flux Data}

Additional photometric data, shown in Tables~\ref{table_more_fluxes_hdfn},
\ref{table_more_fluxes_gss}, and~\ref{table_more_fluxes_mir_radio}, were used
as consistency checks on the absolute flux calibration over the entire
wavelength range. These data were also used to estimate the importance of the
aperture correction when measuring the emission line fluxes (see
Figs.~\ref{figure_sp_hdfn} and~\ref{figure_sp_gss}), and to derive SFRs based
on ultraviolet, infrared and radio luminosities. 

The HST magnitudes for the \mbox{HDF-N} galaxies
(Table~\ref{table_more_fluxes_hdfn}) correspond to the fluxes published by
\citet{fer99}. 

\begin{figure}
\epsscale{0.50}
\plotone{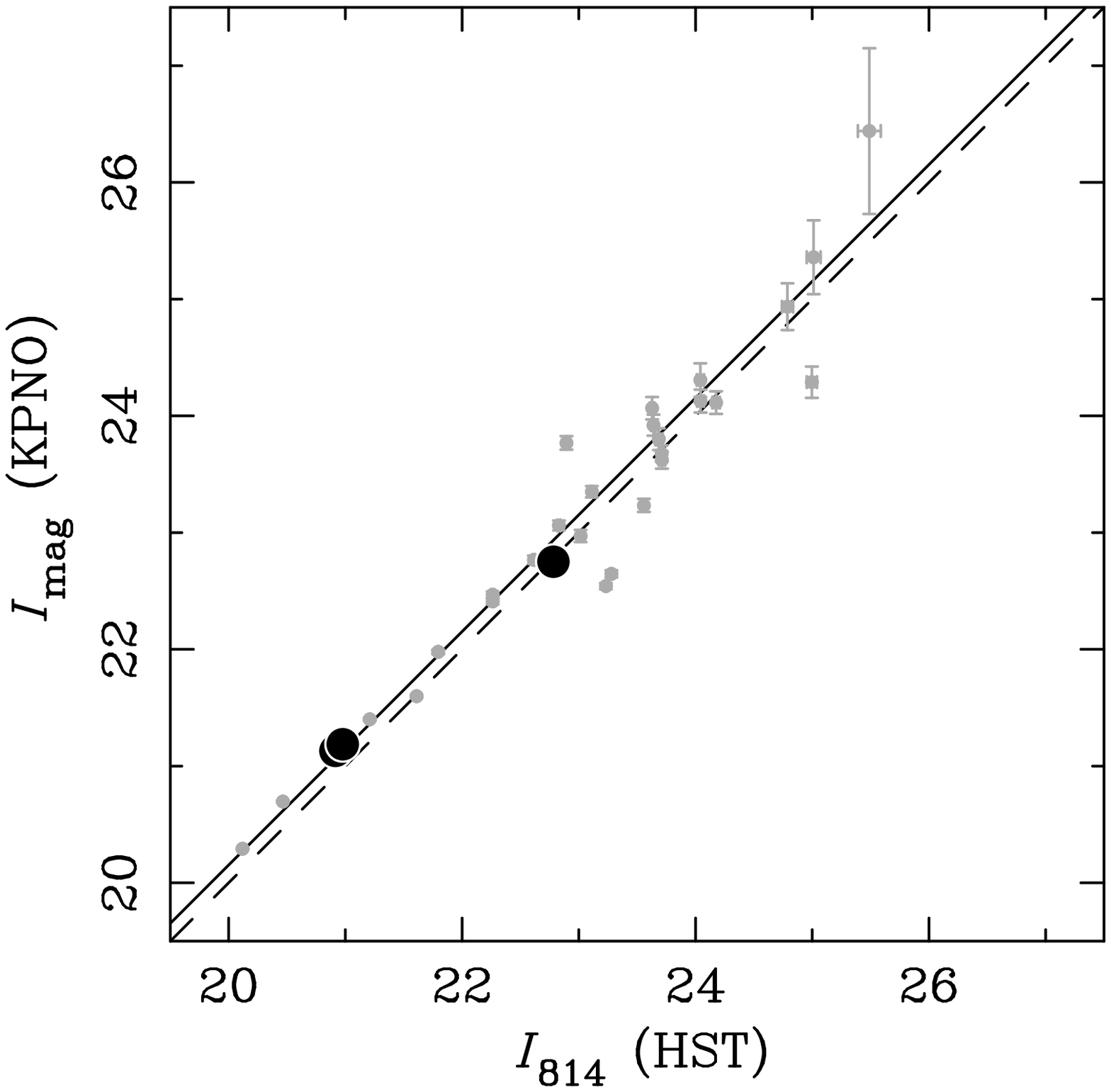}
\figcaption{Comparison of the KPNO $I_{\rm mag}$ with HST $I_{814}$ magnitudes
(AB system).  The data correspond to measurements of objects in field GSS073
(the GSS field with the largest exposure times ---see caption of
Fig.~\ref{figure_mosaic}---).  KPNO magnitudes come from \citet{bru99}. HST
magnitudes have been measured in this work with SExtractor \citep{ber96}. The
three GSS galaxies of our sample are plotted as big circles. The dashed line is
the 1:1 relation, whereas the solid line is the weighted least-squares fit to a
straight line of slope unity.  The offset between both lines is
0.15~magnitudes.\label{figure_plot_brunner}}
\end{figure}

For the GSS galaxies (Table~\ref{table_more_fluxes_gss}) we list the $U$, $B$,
$R$ and $I$ band magnitudes determined by \citet{bru99} using photometric data
obtained with the Prime Focus CCD (PFCCD) camera on the Mayall 4~m telescope at
Kitt Peak National Observatory.  In addition, we have computed $V_{606}$ and
$I_{814}$~band magnitudes using SExtractor \citep{ber96} applied on the HST
images \citep{gro94}.  We have also compared the KPNO magnitudes with their
analogous HST measurements. In this sense, a small offset of 0.15~mag has been
detected between the KPNO $I$ band magnitudes and the HST $I_{814}$ data (see
Fig.~\ref{figure_plot_brunner}). We do not make a similar correction for the
KPNO $R$ band magnitudes because the overlap in wavelength with the $V606$
bandpass is small and the comparison between both magnitudes has a large
scatter.  Finally, we also derived a direct measurement of the $J$ band
magnitude for the pair of interacting galaxies GSS073\_1810. In this case, we
made use of the images obtained with the slit-viewing camera SCAM of NIRSPEC
through the N3 filter (covering the range 1.143-1.375 microns). Although the
total effective exposure time only lasted 80 seconds, it was enough to get the
absolute flux with a relative random uncertainty of $\sim 25$\%.

The MIR flux densities (Table~\ref{table_more_fluxes_mir_radio}),
$S_{\nu}(15\mu{\rm m})$, are the ISOCAM measurements with the LW3 filter
published by \citet{aus99} and \citet{flo99} for the \mbox{HDF-N} and the GSS
galaxies, respectively. Non-detections are indicated by an upper limit, that
corresponds to the sensitivity threshold of the surveys. The flux published by
\citet{aus99} for hd2-264.1 integrates the emission of four blended sources.
The flux value given here is the result for only this galaxy, obtained after
deconvolving the different objects (H.\ Aussel 2001, private communication).
The fluxes for GSS073\_1810 and GSS084\_4521 are also revised values (after
improving the flux calibration through Monte Carlo simulations; H.\
Flores 2001, private communication).

The radio flux densities (last column in
Table~\ref{table_more_fluxes_mir_radio}), $S_{\nu}({\rm radio})$, are the
8.5~GHz measurements from \citet{ric98} for the \mbox{HDF-N} galaxies, and the
5.0~GHz data from \citet{fom91} for the GSS objects.  We also quote the
sensitivity limit of the considered survey for the undetected galaxies.
Finally, to be consistent with the estimation of the MIR flux for hd2-264.1, we
consider that the radio flux in this case also integrates the combined values
of the blended sources. Thus, the quoted radio flux is the same fraction of the
total radio flux.

\section{Constraining the AGN contribution}
\label{agn}
Before calculating the SFRs from the collected multiwavelength data, we must
consider the possible contamination of our galaxy sample by AGNs. The fact that
QSOs and ULIRGs exhibit similar total luminosities and space densities led
\citet{san88} to suggest that ULIRGs may constitute the dust-enshrouded phase
of QSO formation. In this sense, strong interactions and mergers are the
mechanisms responsible for the funneling of gas into the central regions of
galaxies, which subsequently may feed both star formation and AGN activity.
Although some aspects of this scenario are still controversial, observations
confirm the connexion between QSO,
ULIRGs and mergers \citep[][and references therein]{can01}.
In conclusion, rather than segregating LIRGs and ULIRGs
into starbursts and AGNs, the key question is to constrain the relative
contribution of each component to the IR luminosity in a case by case basis.

\subsection{Emission-line diagnostic diagrams}

\begin{figure}
\epsscale{0.80}
\plotone{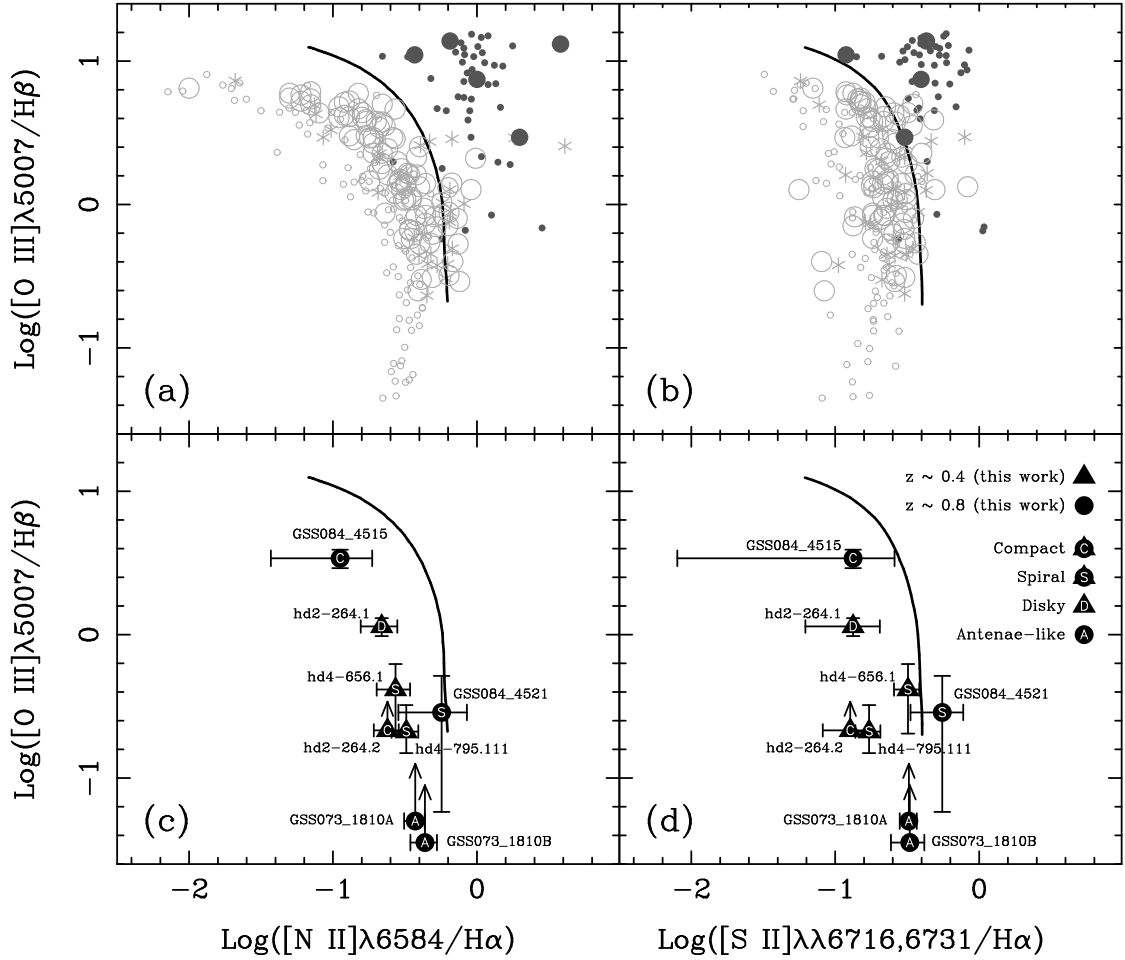}
\figcaption{Diagnostic diagrams for nearby galaxies ---panels (a) and~(b)---
and for our galaxy sample ---panels (c) and~(d)---. {\bf Panels (a) and (b)}:
filled symbols, galaxies with active nuclei; open symbols, galaxies with star
formation; small symbols, galaxies from Figs.~1 and~2 of 
\citet[hereafter VO87]{vei87} (we
only distinguish between emission due to star formation and due to active
nuclei; we have also included the objects classified as NELGs by VO87
---plotted with asterisks---, which are either LINERs or H~{\sc ii}
galaxies); large symbols, UCM galaxies from \citet{gal96} (complete sample of
H$\alpha$ emission-line galaxies at $z < 0.045$); thick solid line: separation
between different type of galaxies. There is a small but clear offset between
the locus of star-forming galaxies in VO87's data, when
compared with the UCM galaxies. This shift is probably caused by the use by
VO87 of data from the literature, which used different slit
apertures. {\bf Panels (c) and (d)}: emission line ratios for our galaxy sample
after applying extinction corrections (triangles for objects 
with $z\sim 0.4$, and
circles for galaxies with $z\sim 0.8$); arrows indicate that the [O~{\sc
iii}]/H$\beta$ ratio is unknown.  In the last two panels we have labeled the
symbols with the letters C, S, D, and A to indicate the morphological type of
each galaxy, as indicated in the symbol key. Since the [O~{\sc
iii}]$\lambda$5007 line for GSS084\_4521 was outside the spectral range of
LRIS, we have estimated its value using the measured [O~{\sc iii}]$\lambda$4959
flux. In the case of the colliding galaxies GSS073\_1810 both lines were not
within the observed spectral interval, so only \mbox{[N~{\sc
ii}]$\lambda$6583/H$\alpha$} and \mbox{[S~{\sc
ii}]$\lambda\lambda$6716,6731/H$\alpha$} are indicated. Both panels indicate
that all the galaxies of our sample fall in the region of star-forming
galaxies.  \label{plot_el}} 
\end{figure}

An important constraint on the contribution of AGN radiation,
if any, to the total luminosity of our galaxies comes from the use of
emission-line diagnostic diagrams. The set of emission lines displayed in
Table~\ref{table_fluxes} allows us to examine two of the most commonly employed
diagrams. In Fig.~\ref{plot_el} we compare the ratio [O~{\sc
iii}]$\lambda$5007/H$\beta$ (which is an indicator of the gas excitation)
versus the ratios [N~{\sc ii}]$\lambda$6583/H$\alpha$ and [S~{\sc
ii}]$\lambda\lambda$6716,6731/H$\alpha$ (the tabulated line ratios are also
given in Table~\ref{table_emission_line_ratios}). Two interesting properties of
these diagrams are their ability to properly discriminate between mechanisms
responsible for the gas ionization (star formation, AGN, or shocks), and their
low sensitivity to extinction corrections. The latter is clearly manifest in
our data, since the mean correction of the plotted ratios in our galaxy sample
is only 2\%, with a maximum of 7\% for the [O~{\sc iii}]$\lambda$5007/H$\beta$
ratio in hd2-264.2. In panels~\ref{plot_el}(a) and~\ref{plot_el}(b) we display
a large collection of emission line ratios corresponding to nearby galaxies
from \citet{vei87} and \citet{gal96} ---see table caption for details---,
whereas in panels~\ref{plot_el}(c) and~\ref{plot_el}(d) we do the same with our
galaxy sample. The analysis of this figure clearly indicates that the locus
spanned by our galaxies is perfectly compatible with that exhibited by star
forming galaxies. 

\begin{figure}
\epsscale{0.80}
\plotone{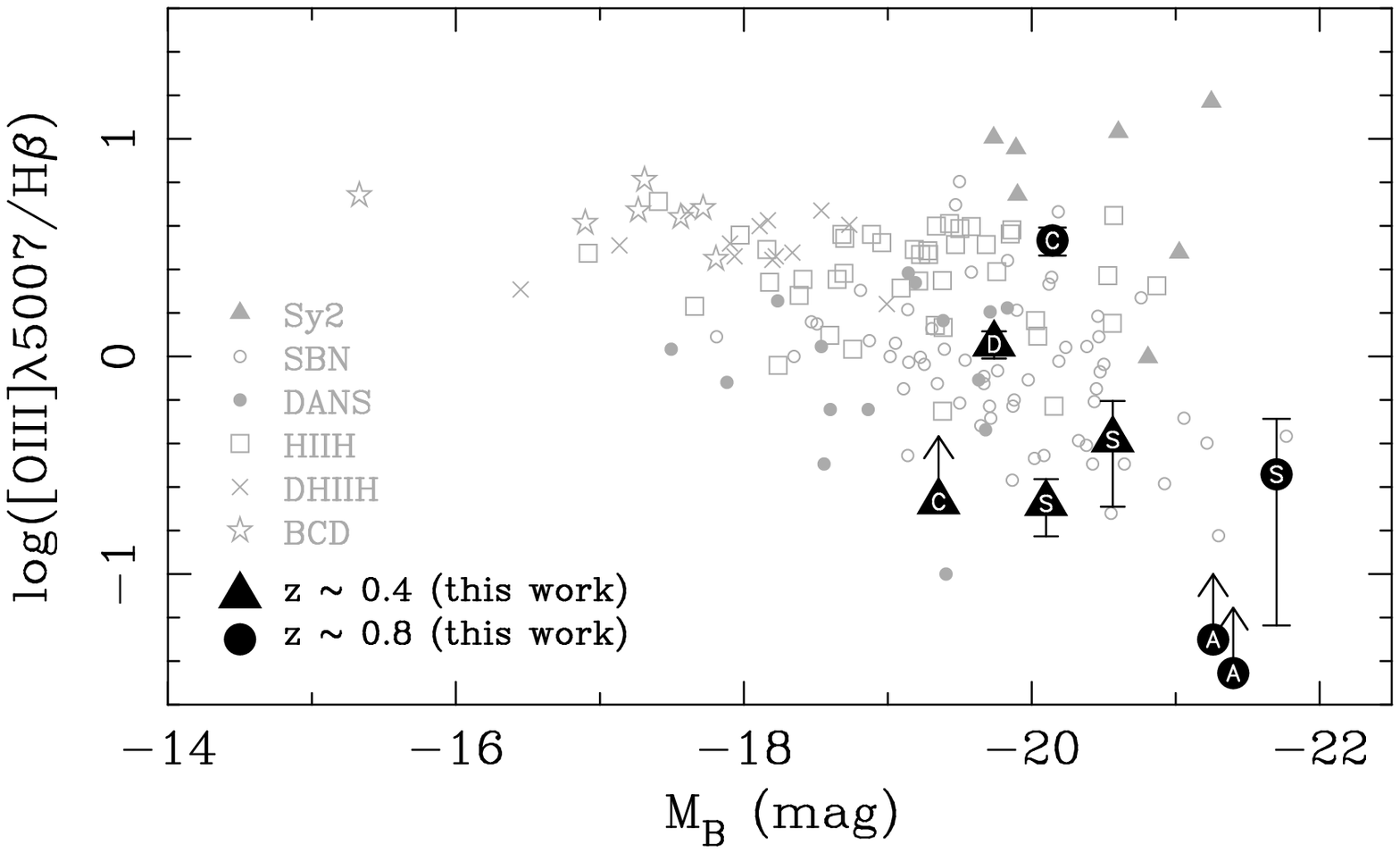}
\figcaption{Excitation as a function of rest-frame absolute blue magnitude for
the galaxies in our sample (large symbols, labeled as in Fig.~\ref{plot_el}),
compared to the sample of nearby UCM ($z < 0.045$) emission-line galaxies of
\citet{gal96,gal97} (small symbols).  The rest-frame absolute $B$ magnitudes
for the UCM sample have been extracted from the data presented by
\citet{per00}, and were corrected for Galactic extinction using the
dust maps of \citet{sch98}.  Different symbols are employed to show the
classification of the local galaxies of different types (see discussion in 
Sect.~\ref{agn}).\label{figure_mbexc}}
\end{figure}

It is also very interesting to compare the measured line fluxes exhibited by
our galaxy sample, and in particular the ratios [O~{\sc
iii}]$\lambda$5007/H$\beta$ (an indicator of the emitting gas excitation), with
those of local star-forming galaxies, as a function of luminosity. For this
purpose, we have estimated rest-frame absolute blue magnitudes for each galaxy
using the available HST $V606$ and $I814$ magnitudes (note that the HST F606W
and F814W bandpasses correspond approximately to the rest-frame $B$~band at
$z\sim 0.4$ and $z\sim 0.8$, which are precisely quite close to the redshifts
exhibited by the present sample). The HST fluxes were corrected for redshift,
the transmission curves of filters F606W and F814W were de-redshifted to
reproduce pseudo $B$ bands for each target, and a residual k-correction
(ranging from $-0.06$ to $0.28$~mag) was derived for each object, using the
averaged spectral energy distributions corresponding to Sab, Sbc and Scd
galaxies of \citet{fuk95}. The resulting magnitudes are listed in
Table~\ref{table_more_galaxy_properties}, and the excitation diagram is
displayed in Fig.~\ref{figure_mbexc}. For comparison, we have included in the
same figure the very well characterized local sample of the UCM survey
\citep{gal96,gal97}, using different symbols to discriminate among distinct
star forming galaxies: SBN (Starburst Nuclei, originally defined by
\citet{bal83}, are spiral galaxies that host a nucleus with an important
star-forming process, with H$\alpha$ luminosities always higher than
10$^8$~L$_\odot$), DANS (Dwarf Amorphous Nucleus Starburst, introduced by
\citet{sal89}, are similar to SBN but at lower scale), HIIH (H~{\sc ii}
galaxies hot spot ---see \citet{gal96}---, are bright galaxies with a global
star-forming process, optical spectrum dominated by blue colors, and strong
emission lines; they show similar H$\alpha$ luminosities to those exhibited by
SBN, but with large [O~{\sc iii}]/H$\beta$ ratios), DHIIH (dwarf H~{\sc ii} hot
spot, are similar to HIIH, but with H$\alpha$ luminosities lower than $5\times
10^7$~L$_\odot$), and BCD (blue compact dwarf, are star forming galaxies with
the lowest luminosity and higher ionization). Finally, Seyfert~2 galaxies (Sy2)
are also displayed in Fig.~\ref{figure_mbexc} to show the plot region covered
by AGN-dominated objects. The analysis of this figure reveals that the galaxies
of our sample are perfectly compatible with the locus of local SBN, with the
exception of hd2-264.2, which based on its higher [O~{\sc iii}]/H$\beta$ ratio,
is located in the region of local HIIH objects. 

\subsection{Further evidence}

There is a growing body of evidence indicating that the population of IR
luminous galaxies detected by the ISOCAM deep surveys are generally
dust-obscured starbursts \citep{fad02,elb02}. \citet{gen98} and \citet{lau00}
report the presence of a dichotomy in the spectroscopic properties of
starbursts and galaxies whose luminosity is dominated by gravitational
accretion: the former show stronger MIR broad emission features, with a fast
decline in their emission shortward of \mbox{$\lambda \sim 6 \mu{\rm m}$},
whereas AGNs are characterized by a flatter spectrum.  Through the analysis of
these features, \citet{tra01} showed that AGN activity is dominant in local
ULIRGs only for very luminous objects.  In particular, these authors found that
the contribution due to star formation to the total far-infrared luminosity is,
on average, 82\%--94\% for galaxies with $L_{\rm IR} < 10^{12.4} {\rm
L}_{\odot}$ (for a cosmology with $H_0=75 \; {\rm km} \; {\rm s}^{-1} \; {\rm
Mpc}^{-1}$, and  $q_0=0.5$), these numbers decrease to 44\%--55\% at higher
luminosities.  Considering that all our galaxies are below that threshold (see
Table~\ref{table_more_galaxy_properties}) our data support the idea that star
formation is the dominant source of ionization in our sample (note that the
largest $L_{\rm IR}$ value of $\sim$ 10$^{12}$ L$_{\odot}$ corresponds to
GSS073\_1810, which actually consists of two interacting galaxies).

Since accretion around black holes must lead to the formation of luminous X-ray
sources, specially in the hard X-ray domain (less sensitive to extinction
affecting obscured AGNs), the cross-correlation of deep X-ray and MIR
observations is a direct way to check for the mechanism responsible for the
nebular emission-lines. From the deep X-ray survey in the HDF-N performed with
Chandra \citep{hor00,hor01,bra01}, we find that only one of our four galaxies
in this field, hd2-264.1, is an X-ray source (see Fig.~5 in Brandt et al.\
2001). Very recently \citet{fad02} have studied in detail the mechanisms that
power the X-ray emission of ISOCAM galaxies in the HDF-N+FF presenting X-ray
counterparts. They found that hd2-264.1 exhibits an X-ray luminosity $L_{\rm X}
\sim 2 \times 10^{41} {\rm erg\;s}^{-1}$ and an X-ray spectral index similar to
those of the star forming galaxies Arp~220 and Arp~244. Although we are not
going to repeat all the details here, it is important to stress that the latter
result alone clearly suggests that, in spite of its X-ray detection, a dominant
AGN contribution in this object can be safely dismissed.

\begin{figure}
\epsscale{1.00}
\plotone{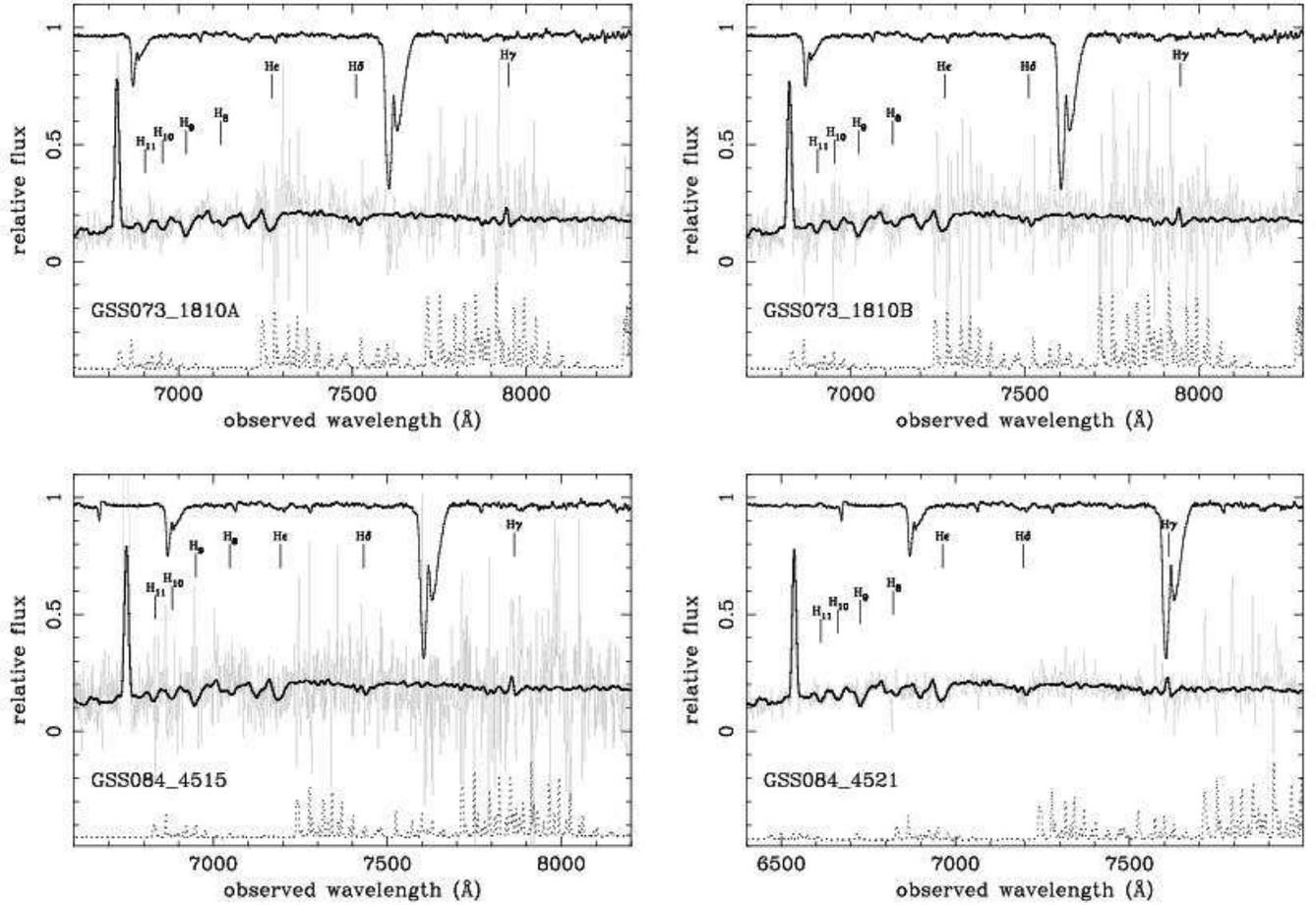}
\figcaption{Comparison of the GSS subsample spectra with a local Sc galaxy
spectrum. The thin line (light gray) is the spectrum of each galaxy (in
arbitrary units); the upper spectrum is the normalized atmospheric
transmission; the dotted line is the corresponding sky spectrum (in arbitrary
units and shifted downwards); the thick line is the spectrum of NGC~4775, a
typical Sc galaxie from the spectrophotometric atlas of \citet{ken92b},
redshifted and scaled to match the continuum level of each galaxy spectrum. The
wavelength location of the Balmer lines present in the plotted spectral range
are also shown. Note that, in spite of the unavoidable large residuals
associated to bright sky lines, there is good agreement between the GSS
spectra and the scaled Sc spectrum. Since in all the panels the SED of NGC~4775
is plotted with the same scale, the different strength of [O~{\sc
ii}]$\lambda$3727 reveals that the GSS galaxies span a range in the equivalent
width of this emission line.  \label{figure_balmer}}

\end{figure}

Further support to the idea that the optical light of the galaxies under study
is dominanted by hot stars is given by the LRIS spectra of the GSS subsample.
These have enough S/N that a cross-correlation of their spectral energy
distribution (SED) against well-known SEDs of local star forming galaxies is
easily done.  This task was carried out by fitting the continuum regions free
of high sky-line residuals to the SEDs in the spectrophotometric atlas of
\citet{ken92b} masking out emission line regions. One of the best fits is
obtained with NGC~4775, a typical Sc galaxy. This result, shown in
Fig.~\ref{figure_balmer}, reveals that typical high-order absorption Balmer
lines, a signature of young stellar populations, are present in the spectra of
these targets.

A final piece of data gives more support to the same idea:
the measured emission-line widths are not broad. The averaged
FWHM values (derived from the fits to the emission lines with better S/N and
without high residuals of sky lines) after quadratically subtracting the
spectral resolution, are (in km/sec): $110 \pm 9$ (hd2-264.1), $94 \pm 9$
(hd2-264.2), $101 \pm 7$ (hd4-656.1), $129 \pm 9$ (hd4-795.111), $216 \pm 26$
(GSS073\_1810A), $167 \pm 45$ (GSS073\_1810B), $327 \pm 42$ (GSS084\_4515), and
$216 \pm 35$ (GSS084\_4521). With this result, we are confident that none of
the galaxies under study is a Seyfert~2.

In conclusion, although we cannot completely rule out AGNs as a possible source
of ionizing radiation, all evidence compiled here suggests that such a
contribution, if present, is very small.

\section{Multiwavelength star formation rates}
\label{sfr}
We have transformed the previous flux measurements and their corresponding
errors (Tables~\ref{table_fluxes}, \ref{table_more_fluxes_hdfn}
and~\ref{table_more_fluxes_gss}) into star formation rates, using well-known
calibrations. The summary of this computation is presented in
Table~\ref{table_sfrs}, displayed in Fig.~\ref{figure_sfrs}, and described in
this section.

\subsection{SFR from ultraviolet continua} 

Since the resulting UV spectrum of star forming galaxies is roughly
flat in the range of 1500--2800~\AA, the SFR can be derived from any
measurement of the UV luminosity in that spectral interval, using a
calibration such as that given in Eq.~(1) of \citet{ken98}, namely,
\begin{equation}
{\rm SFR}_{\rm UV}(M_\odot/{\rm yr}) = 1.4 \times 10^{-28} \; L_{\rm UV}
 ({\rm erg} \; {\rm s}^{-1} \; {\rm Hz}^{-1}).
\label{equation_sfruv}
\end{equation}
Note that at the mean redshifts of the \mbox{HDF-N} and GSS galaxies of our
sample, the $U$~band magnitudes displayed in
Tables~\ref{table_more_fluxes_hdfn} and~\ref{table_more_fluxes_gss} ---the HST
F300W ($\langle\lambda\rangle \sim 3000$~\AA) and the KPNO $U$
($\langle\lambda\rangle\sim 3670$~\AA) filters---  sample practically the same
rest-frame wavelength $\lambda_{\rm rest}\sim 2050$~\AA. The estimated
luminosities were corrected for internal reddening using the recipe of
\citet{cal00} for starbursts, where the color excesses for the stellar
continuum were derived using E(B$-$V)$_{\rm cont}=0.44\;$E(B$-$V)$_{\rm gas}$
\citep{cal97}. In addition, the Galactic extinction has been corrected using
the E$_{\rm G}$(B$-$V) from Table~\ref{table_sample} and the extinction curve
of \citet{fit99}.

The errors were determined numerically by bootstrapping the original spectra
and their corresponding error spectra. Thus, the bootstrap propagates the
errors in E(B$-$V)$_{\rm gas}$ quoted in
Table~\ref{table_fluxes}.  However, we have not included here the intrinsic
scatter in Eq.~(\ref{equation_sfruv}). In fact, different calibrations between
UV flux and SFR differ by $\sim 0.3$~dex \citep{ken98}. In addition, systematic
errors may also be present due to the fact that Eq.~(\ref{equation_sfruv}) is
valid only for galaxies with continuous star formation over time scales of
$10^{8}$ or longer.

\subsection{SFR from {\rm [O~{\sc ii}]} and {\rm H$\alpha$} luminosities}

Here we use the average calibrations given in Eq.~(3) and~(2) of
\citet{ken98},
\begin{equation}
{\rm SFR}_{\rm [O~{\mbox{\sc ii}}]}(M_\odot/{\rm yr}) =
 1.4 \times 10^{-41} \; L_{\rm [O~{\mbox{\sc ii}}]}({\rm erg}\;{\rm s}^{-1}), 
\label{equation_sfroii}
\end{equation}
and
\begin{equation}
{\rm SFR}_{{\rm H}\alpha}(M_\odot/{\rm yr}) =
 7.9 \times 10^{-42} \; L_{{\rm H}\alpha}({\rm erg}\;{\rm s}^{-1}), 
\label{equation_sfrhalpha}
\end{equation}
with the correction for internal reddening was applied as described above.  It
is important to keep in mind that $L_{{\rm H}\alpha}$ traces the instantaneous
SFR, because the H$\alpha$ emission is due to the re-processing of ionizing
radiation shortward of 912~\AA, which is produced by the most massive stars.
Thus they are strongly dependent on the adopted initial mass function.

Since the errors in E(B$-$V)$_{\rm gas}$ and the extinction-corrected H$\alpha$
flux are correlated, the error spectra were also bootstrapped to calculate the
errors for SFR$_{{\rm H}\alpha}$. We should also note that
in this procedure we have not included the intrinsic scatter in
Eqs.~(\ref{equation_sfroii})--(\ref{equation_sfrhalpha}).

Finally the values derived for ${\rm SFR}_{\rm [O~{\mbox{\sc ii}}]}$ and ${\rm
SFR}_{{\rm H}\alpha}$, and their corresponding error bars, were corrected for
aperture effects. For that purpose, we used the factors listed in
Table~\ref{table_aperture_corrections} for the bluest filters. In particular,
\mbox{HDF-N} galaxies were corrected using the fraction of galaxy light passing
through the $U_{300}$ filter, whereas for the GSS objects we employed the
factors associated to the $V_{606}$ filter (slit 1.18\arcsec\ for ${\rm
SFR}_{\rm [O~{\mbox{\sc ii}}]}$; slit 0.76\arcsec\ for ${\rm SFR}_{{\rm
H}\alpha}$).  These aperture corrected SFRs are the ones tabulated in
Table~\ref{table_sfrs}.

\subsection{SFR from mid infrared fluxes}
\label{subsection_sfr_ir}

\begin{figure}
\epsscale{0.90}
\plotone{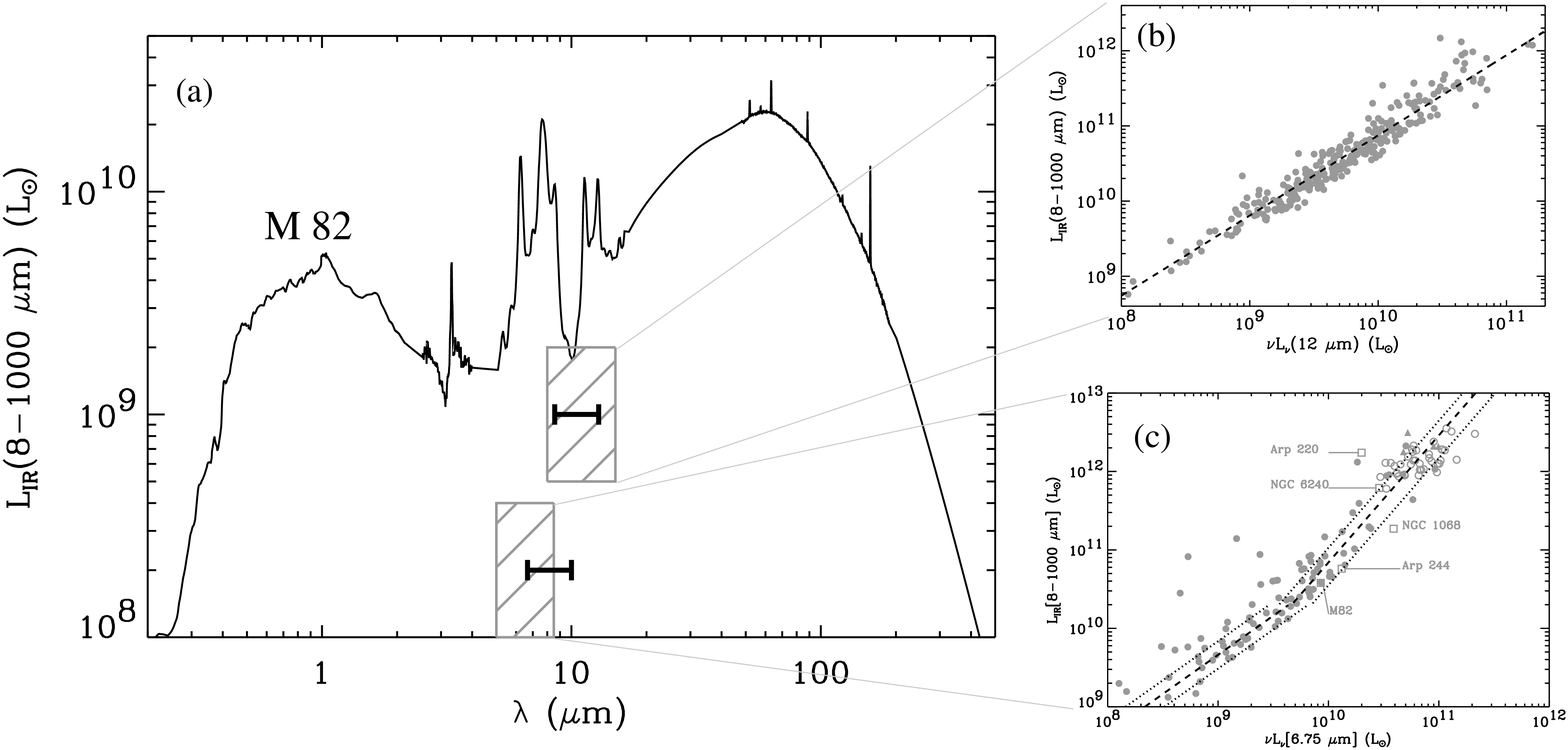}
\figcaption{{\bf Panel~(a)}: SED of the proto-typical starburst M 82 \citep[as
in][]{elb02}.  The two thick line segments mark the width and position of
the ISOCAM-LW3 filter if the galaxy were located at $z=$ 0.4 (upper one) and
$z=$ 0.8 (lower one). The dashed rectangular regions mark the position and
width of the IRAS-12\,$\mu$m filter (upper one) and ISOCAM-LW2 filter (lower
one). {\bf Panel~(b)}: Integrated IR luminosity, $L_{\rm IR}$, versus
IRAS-12\,$\mu$m monochromatic luminosity in $\nu L_{\nu}$ for 293 IRAS BGS
galaxies. {\bf Panel~(c)}: Integrated IR luminosity, $L_{\rm IR}$, versus
ISOCAM-LW2 (6.75\,$\mu$m) monochromatic luminosity in $\nu L_{\nu}$. This
figure, which contains 91 galaxies, comes from Fig.~5d of \citet{elb02}.
\label{figure_m82}}
\end{figure}

The SFR of a galaxy may be estimated from its integrated IR
luminosity, $L_{\rm IR} = L(8-1000\,\mu{\rm m})$ using Eq.~(4) of
\citet{ken98},
\begin{equation}
{\rm SFR}_{\rm IR}(M_\odot/{\rm yr}) = 
 4.5 \times 10^{-44} \; L_{\rm IR}({\rm erg}\;{\rm s}^{-1}).
\label{equation_sfr_ir}
\end{equation}
The IR SED of local star forming galaxies generally peaks in the FIR,
between $\lambda \sim$ 40 and 100\,$\mu$m. In this spectral region,
most of the luminosity is due to the thermal emission of the so-called
big dust grains. The MIR flux is a combination of broad emission
features and hot dust continuum as in the SED of the proto-typical
starburst galaxy M 82 presented in Fig.~\ref{figure_m82}a. Our sample
of galaxies was observed in the LW3 filter of ISOCAM, centered at
15\,$\mu$m ($\lambda=$ 12--18\,$\mu$m).

At the redshift of the HDF-N galaxies ($z\sim$ 0.4), the ISOCAM-LW3
filter (upper thick horizontal segment on Fig.~\ref{figure_m82}a)
samples almost the same rest-frame wavelength as the IRAS-12\,$\mu$m
filter for galaxies at $z$= 0. In Fig.~\ref{figure_m82}b, we show that
the 12\,$\mu$m luminosity ($\nu L_{\nu}$) of local galaxies is tightly
correlated with their integrated IR luminosity, $L_{\rm IR}$. We used
the 293 galaxies from the IRAS Bright Galaxy Sample
\citep[BGS][]{soi89} with 12\,$\mu$m flux densities and redshifts to
derive this correlation.

At the redshift of the GSS galaxies ($z\sim$ 0.8), the ISOCAM-LW3
filter (lower thick horizontal segment on Fig.~\ref{figure_m82}a)
largely overlaps the same rest-frame wavelength as the ISOCAM-LW2
filter (upper dashed rectangle), which is centered at 6.75\,$\mu$m
($\lambda=$ 5--8.5\,$\mu$m). In Fig.~\ref{figure_m82}c, we have
reproduced the Fig.~5d from \citet{elb02} which shows that rest-frame
LW2 luminosities are also correlated with $L_{\rm IR}$.

The previous two correlations suggest that we can compute $L_{\rm IR}$
for galaxies with $z\sim$ 0.4 and 0.8 which were detected at
15\,$\mu$m. However, because the rest-frame wavelength range
corresponding to the observed 15\,$\mu$m ISOCAM band does not exactly
match the ISOCAM-LW2 and IRAS-12\,$\mu$m filters, we did not compute
$L_{\rm IR}$ directly from the correlations but instead from a library
of template SEDs ranging from 0.1 to 1000~$\mu{\rm m}$. These
templates were produced by \citet{cha02} who combined observed SEDs in
order to reproduce the same observed trends between mid-infrared and
far-infrared luminosities derived from ISOCAM, IRAS and SCUBA
observations of nearby galaxies.

This computation relies on the hypothesis that the trends observed for
local galaxies apply also to galaxies located at $z\sim$ 0.4 and
0.8. As suggested in \citet{elb02}, this assumption is supported by
the correlation observed between radio and MIR luminosities both in
the local and distant universe ($z\sim$ 0.8).

To obtain the total integrated IR luminosity for the galaxies of our
sample, we proceeded as follows. We computed the expected flux density,
$S_\nu(15\mu{\rm m})$, for all the template SEDs, at the redshifts
of each of our sample galaxies. The values thus obtained were compared
to the measured ISOCAM-15\,$\mu$m flux densities in order to select,
for each sample galaxy, the template SED that best matched its ISOCAM
flux density. The chosen SED was then used to predict IRAS flux
densities. The latter were employed to compute the total IR flux densities,
using
\begin{equation}
\label{EQ:Lir}
\frac{F_{\rm IR}(L_\odot)}{1.8 \times 10^{-14}} =
       13.48 \; f_{12} + 5.16 \; f_{25} + 2.58 \; f_{60} + f_{100},
\end{equation}
\citep{san96}, where $f_{12}$, $f_{25}$, $f_{60}$ and $f_{100}$ are the
IRAS flux densities in Jy at 12, 25, 60 and 100 $\mu m$.

Note that we did not compute $L_{\rm IR}$ by directly integrating the
template SED because the spectral energy distributions of
\citet{cha02} were built to fit correlations using Eq.~\ref{EQ:Lir}.
Anyway, the difference in the computation of $L_{\rm IR}$ between
using Eq.~\ref{EQ:Lir} and integrating the SED is of the order of only
5--10\%. The final IR luminosity for each galaxy is listed in
Table~\ref{table_more_galaxy_properties}.

The error bars in ${\rm SFR}_{\rm IR}$ quoted in
Table~\ref{table_sfrs} have been computed by propagating the errors in
the ISOCAM-15\,$\mu$m flux (i.e.\ obtaining the best SED which fitted
$S_\nu(15\,\mu{\rm m}) \pm {\rm error}$). They do not include a
possible intrinsic scatter within the correlations shown in the
Figs.~\ref{figure_m82}b,c since the observed scatter combines this intrinsic
scatter with measurement errors. If the scatter observed in 
Figs.~\ref{figure_m82}b,c were mostly intrinsic, the 1-sigma error bars on
${\rm SFR}_{\rm IR}$ quoted in Table~\ref{table_sfrs} should be increased
according to a 40\% uncertainty on $L_{\rm IR}$.

\subsection{SFR from radio fluxes}
\label{sfr_from_radio}

The correlation between radio and infrared luminosities
\citep{hel85,con91,yun01} offers an additional diagnostic tool to estimate the
SFR.  This correlation can be characterized with the help of the flux density
ratio $q$ defined by \citet{hel85},
\begin{equation}
q = \log\left(
  \frac{F(\mbox{40--120}\;\mu{\rm m})}{3.75\times 10^{12}\;{\rm W\;m}^{-2}}
        \right) 
   -\log\left(
  \frac{S_\nu(1.4\;{\rm GHz})}{{\rm W\;m}^{-2}\;{\rm Hz}^{-1}}
        \right),
\label{equation_q_parameter}
\end{equation}
where $S_\nu(1.4\;{\rm GHz})$ is the observed 1.4~GHz flux density in units of
${\rm W\;m}^{-2}\;{\rm Hz}^{-1}$. In the study of a complete sample of
1809 galaxies from the IRAS 2~Jy sample, \citet{yun01} have shown that 98\% of
the galaxies follow the above relation with $q = 2.34 \pm 0.01$, the scatter in
the linear relation being $\sim 0.26$~dex.

We computed rest-frame radio luminosities at 1.4~GHz assuming that $S_\nu
\propto \nu^{-\alpha}$ \citep[as suggested by][]{yun01}, using the radio fluxes
given in Table~\ref{table_more_fluxes_mir_radio}, and the typical radio
spectral index $\alpha = 0.80 \pm 0.15$, with the exception of hd2-264.1
($\alpha=0.87\pm 0.12$) and GSS084\_4521 ($\alpha=0.75\pm 0.25$), whose values
were determined more precisely due to the availability of radio measurements at
two different frequencies.

To convert the derived FIR luminosity, $L(\mbox{40--120}\,mu{\rm m})$,
into integrated IR luminosity, $L(\mbox{8--1000}\,mu{\rm m})$, we used the
following tight correlations derived from the IRAS BGS \citep{soi89}
\begin{equation}
L(\mbox{8--1000}\;\mu{\rm m}) = (1.26 \pm 0.15) \; 
                                 L(\mbox{40--500}\;\mu{\rm m}),
\end{equation}
\begin{equation}
L(\mbox{40--500}\;\mu{\rm m}) = (1.50 \pm 0.10) \; 
                                 L(\mbox{40--120}\;\mu{\rm m}),
\end{equation}
and thus
\begin{equation}
L(\mbox{8--1000}\;\mu{\rm m}) = (1.89 \pm 0.26) \; 
                                 L(\mbox{40--120}\;\mu{\rm m}).
\label{equation_fir_mir}
\end{equation}
Finally, the SFR$_{\rm radio}$ is computed from Eq.~(\ref{equation_sfr_ir}),
employing the $L(\mbox{8--1000}\;\mu{\rm m})$ luminosity obtained from the
combination of Eqs.~(\ref{equation_q_parameter}) and~(\ref{equation_fir_mir}).

The random errors correspond to the coadded effect of random errors in the
radio flux and in the spectral index $\alpha$, neglecting the contribution of a
possible intrinsic scatter in the $q$ parameter.

\section{Additional galaxy properties}
\label{galaxy_properties}

Although the galaxy sample examined in this paper is small, it is clear from
Fig.~\ref{figure_mosaic} that it includes a diverse collection of objects. In
addition to star formation rate estimates, the sample can also be used to
derive additional galaxy properties, like metallicities and masses, which will
provide a more complete description of each object.

Metallicities can be determined from the measured emission line strengths using
the empirical line flux ratio technique of \citet{pag79}, which relates the
line ratio 
\begin{equation}
{\rm R}_{23}= \frac{\mbox{[O~{\sc ii}]}\lambda3727+
 \mbox{[O~{\sc iii}]}\lambda\lambda4959,5007}{{\rm H}\beta}
\end{equation}
to the oxygen abundance, O/H. This calibration, shown in
Fig.~\ref{figure_metallicities}a, 
can be parameterized as a function of the ionization parameter
\begin{equation}
{\rm O}_{32}= \frac{\mbox{[O~{\sc iii}]}\lambda\lambda4959,5007}%
{\mbox{[O~{\sc ii}]}\lambda3727}.
\end{equation}
In this work we have adopted the analytic fits of \citet{kob99} ---in
particular their Eqs.~(7)--(9)--- to the set of photoionization models of
\citet{mcg91}. As seen in Fig.~\ref{figure_metallicities}a, these
calibrations are degenerate, there being two widely different oxygen 
abundances for
a given pair of R$_{23}$ and O$_{32}$ values. Fortunately the diagnostic
diagram relating [N~{\sc ii}]$\lambda6583$/[O~{\sc ii}]$\lambda3727$ with
[N~{\sc ii}]$\lambda6583$/H$\alpha$ (see Fig.~\ref{figure_metallicities}b), has
proven to be an excellent tool to break this degeneracy, as shown by
\citet{con02}. In particular, galaxies that fall in the upper right quadrant of
this diagram can be safely assigned to the upper branch (high metallicity) of
the oxygen calibration in Fig.~\ref{figure_metallicities}a.  This is the case
for all galaxies in our sample, excepting GSS084\_4515, which falls in the
turnaround region of the calibration diagram. Even though the measures for this
galaxy present large uncertainties, it is clear that it exhibits the lowest
metallicity.  We have also included in Figs.~\ref{figure_metallicities}a
and~\ref{figure_metallicities}b the UV selected galaxies from \citet{con02} to
illustrate the reliability of this method. 
The R$_{23}$ and O$_{32}$ metallicity indicators, together
with the final derived metallicities, are listed in
Table~\ref{table_emission_line_ratios}. Note that for the two colliding
galaxies GSS073\_1810 the [O~{\sc iii}]$\lambda\lambda$4959,5007 emission lines
fall, unfortunately, outside the observable wavelength range of LRIS, so we
cannot derive their metallicity.  However, the location of these two objects in
Fig.~\ref{figure_metallicities}b (shown as open circles), suggests that they
also should have a metallicity corresponding to the upper branch of the
R$_{23}$ calibration in Fig.~\ref{figure_metallicities}a.  This result is
fully consistent with the position of our galaxies in the excitation
diagram presented in Fig.~\ref{figure_mbexc}, in which the local starburst
trend is typically interpreted as a sequence governed by the metallicity
content of the ionized gas, with metallicity increasing when moving from the
H~{\sc ii} galaxies to the SBNs.

The locus of our galaxy sample in the metallicity-luminosity diagram is
examined in Fig.~\ref{figure_metlumi}, and compared with the regions
spanned by additional samples of galaxies, extracted from Fig.~10a of
\citet[][see references therein]{con02}. In this sense, hd2-264.1, hd4-656.1,
hd4-795.111, and GSS084\_4521 are perfectly compatible with the locus spanned
by local SBN galaxies (open circles).  In addition, GSS084\_4515 overlaps with
the luminous compact galaxies (LCGs) of \citet{ham01}, whereas hd2-264.2 is
close to the location of the $z \sim \mbox{0.1--0.5}$ emission-line objects of
\citet{koz99}. Finally, since the positions of the two interacting galaxies
GSS073\_1810 in Fig.~\ref{figure_metlumi} indicate that their metallicities
are likely to be in the range covered by the remaining
spiral galaxies of the sample, and considering their rest-frame absolute $B$
magnitude, their location in Fig.~\ref{figure_metlumi} should also be
compatible with the SBN objects.

Galaxy masses for the HDF-N subsample have been estimated using the 
$K$~band photometry
published by \citet{fer99} and listed in Table~\ref{table_more_fluxes_hdfn}.
Using the spectral energy distribution of late-type galaxies given by
\citet{fuk95}, we have derived an average K-correction of
$K(z)-K(0)=-0.65$~mag. Finally, we have employed a mass-to-light ratio
0.93~M$_\odot$/L$_{K,\odot}$, which is the value for the local SBN-like
galaxies modeled by \citet{gil00}. The resulting masses are listed in
Table~\ref{table_more_galaxy_properties}.

We also quote in Table~\ref{table_more_galaxy_properties} values for the galaxy
half-light radius, which come from the Medium Deep Survey \citep{rat99} fits in
the F606W filter for HDF-N galaxies (roughly rest-frame $B$ band at $z\sim
0.4$), or from GIM2D \citep{sim02} fits to the F814W filter (close to the
rest-frame $B$ band at $z\sim 0.8$) in the case of GSS galaxies. The
uncertainties for the HDF-N data were estimated from the variation of the
half-light radius values when using the F300W and F814W filters. The errors for
the GSS galaxies combine the uncertainties calculated by GIM2D using Monte
Carlo simulations, added to the variation in the galaxy fits using the F606W
images.

We have combined the computed rest-frame absolute $B$ magnitudes, with the
half-light radii and the emission-line widths, to compare the structural
properties of our galaxy sample (see Fig.~\ref{figure_global}) with those of
different galaxy types (both local and at redshifts $\lesssim 1$) compiled and
published by \citet{phi97}. The unambiguous spiral-type galaxies of our sample
(hd4-656.1, hd4-795.111, GSS073\_1810 and GSS084\_4521) fall in the region
spanned by local galaxies of this type in these figures (since these objects
are observed face-on, their positions in Fig.~\ref{figure_global}c are only
lower limits).  The disky galaxy hd2-264.1 is also compatible with low
luminosity and low mass spirals, whereas the compact object hd2-264.2 is
consistent with the CNELG and compact galaxies at higher redshifts of
\citet{phi97} and \citet{guz97}.  Interestingly, another compact galaxy from
our sample, GSS084\_4515, behaves in a rather different fashion when plotted in
these diagrams. The latter result is examined in more depth in the next
section.

\section{Description of individual galaxies}
\label{description_individual_galaxies}

Here we present a summary of the properties of each galaxy in our sample.

\subsection{hd2-264.1}
This galaxy appears to be closely connected with hd2-264.2 (the difference in
radial velocity between both objects is only $\sim 350$~km/s), and the star
formation episode may well have been triggered as the result of the interaction
between both galaxies. In the HST images hd2-264.1 shows a central blue
condensation and a close to edge-on diffuse disk. Based on the standard
emission-line ratio diagnostics, this object qualifies as a starburst nucleus
of moderate luminosity (M$_{\rm B}=-19.7$), similar in H$\alpha$ luminosity,
far-IR luminosity, color and size to the local galaxies in the same
spectroscopic category.

\subsection{hd2-264.2}
This object is a featureless compact galaxy close to hd2-264.1. Its appearance
in the HST images is highly concentrated, with a half-light radius of $\sim
0.6$~kpc in the F300W band, and $\sim 1.5$~kpc in the F814W band. Its
low luminosity and small velocity dispersion indicate a low mass system where a
star-forming process, likely motivated by an interaction with its larger
companion hd2-264.1, has increased its luminosity. Its spectrum confirms those
characteristics typical of a high ionization ongoing star formation process.
The object was not detected by ISO, and shows similar properties to the compact
star-forming galaxies at intermediate redshifts studied by different authors
\citep{phi97,guz97,koz99}.

\subsection{hd4-656.1}
This face-on spiral galaxy exhibits two well defined spiral arms in the
HST images. The object shows emission lines that correspond to a moderate
starburst in the nucleus. The galaxy properties of this object are similar to
those of its local counterparts.

\subsection{hd4-795.111}
This galaxy has very similar properties, within errors, to those of
hd4-656.1.  Both show moderate infrared luminosities and spectral features
consistent with a nuclear starburst, with a star formation rate of
about 2 M$_\odot$~yr$^{-1}$ as indicated by all the SFR tracers.

\subsection{GSS073\_1810}
The HST images of GSS073\_1810 reveal a system formed by two morphologically
disrupted spiral galaxies ($\Delta v_{\rm r} \sim 70$~km~s$^{-1}$ in
H$\alpha$). This large ($\sim 27$~kpc total diameter) and luminous system is
clearly undergoing a strong burst of star formation. It is worth noting that
each individual component could be considered as being average in size and
luminosity. The combined system, based on its integrated IR luminosity, can
barely be qualified as an ultraluminous infrared galaxy (ULIRG, log L$_{\rm
IR}$/L$_{\odot}$=11.97). The ratio SFR$_{\rm IR}$ over SFR$_{\rm H\alpha}$ is
well above the $\sim$1 observed for objects with IR luminosity lower than
10$^{11}$L$_\odot$, confirming a different nature for the object.

\subsection{GSS084\_4515}
The behavior of this interesting galaxy differs significantly from hd2-264.2,
the other compact object analyzed in this work. It shows the lowest metallicity
of the galaxy sample, and its value matches that of the sample of LCGs at $0.5
\la z \la 0.7$ (asterisks in Fig.~\ref{figure_metlumi}) of \citet{ham01}.  Its
mass ($\sim 4\times 10^{10}$~M$_\odot$, as read from
Fig.~\ref{figure_global}c), is also similar to that of the
LCG sample (\mbox{$10^{10}$--$2.5\times 10^{11}$~M$_\odot$}).  GSS084\_4515 has
an amorphous but compact morphology (half-light radius $\sim 2.5$~kpc), and its
spectrum is dominated by a high ionization starburst that accounts for the
HII-like emission-line ratios. The colors, SFRs, diagnostic diagrams, H$\alpha$
luminosity, absolute magnitude and metallicity, all suggest that this galaxy is
similar to the brightest compact star-forming galaxies studied by
\citet{guz97}, in particular the subsample at $z>0.7$ classified as disk-like.

\subsection{GSS084\_4521}
This object has the appearance of a bright face-on spiral galaxy, and its
spectroscopic properties are typical of systems presenting a nuclear starburst.
The observed discrepancies between SFR$_{\rm IR}$ and SFR$_{\rm radio}$ with
SFR$_{\rm UV}$, SFR$_{\mbox{[O~{\sc ii}]}}$ and SFR$_{{\rm H}\alpha}$ suggest
that its nuclear starburst is highly obscured. The IR luminosity is very high
($\log{\rm L}_{\rm IR}=11.59$) but similar to extreme local objects such as
UCM2250+2427.

\section{Comparison of different SFR estimators}
\label{sfr_comparison}

Although it can be argued that the calibrations adopted in Sect.~\ref{sfr} to
derive the SFR estimates listed in Table~\ref{table_sfrs} do suffer from large
uncertainties, in this work we are precisely seeking systematic
deviations when using the same prescriptions in all the galaxies. The
discrepancies found by following this approach must then be understood in terms
of processes that violate the hypothesis of a similar star forming scenario in
all the galaxies. Among these, varying star formation histories and differences
in the wavelength dependent extinction corrections are expected to be two of
the most important factors. In addition, the detailed modeling of nebular
emission from star-forming objects by \citet{cha01} shows that the situation
is even more complex since the zero-age effective ionization parameter, the gas
metallicity and the effective dust-to-heavy element ratio are all parameters
which variation introduces scatter in the predicted SFRs.

Even assuming the reduced size of our galaxy sample, the graphic comparison
presented in Fig.~\ref{figure_sfrs} does provide an interesting insight
concerning how different star formation indicators behave in galaxies at
intermediate redshifts.

\subsection{SFR$_{\rm UV}$ and SFR$_{{\rm H}\alpha}$}

The comparison between SFR$_{\rm UV}$ and SFR$_{{\rm H}\alpha}$
(Fig.~\ref{figure_sfrs}a), shows that for galaxies at $z \sim 0.4$, excluding
hd2-264.1, the SFR$_{\rm UV}$ values are slightly above the 1:1 relation. In
the specific case of hd2-264.1, the HST images suggest that we are observing an
almost edge-on disk, so the UV extinction correction might be underestimated
and the actual UV flux be slightly higher.
Interestingly, all the galaxies at higher redshift ($z \sim 0.8$) exhibit
underestimated SFR$_{\rm UV}$. These results are consistent with the work of
\citet{sul01}, who found that the UV/H$\alpha$ ratio decreases with increasing
SFR for galaxies in the redshift range $0 < z < 0.5$, whereas for low-SFR
objects the UV luminosities lead to higher SFRs than H$\alpha$. Previous
observations of distant galaxies, using uncorrected H$\alpha$ fluxes, had
already revealed that the UV flux underestimates the SFR by a factor of a few
\citep{gla99,yan99,moo00}. The origin of this dichotomic behavior has
tentatively been attributed to temporal variations in the star formation
histories, with episodic and rapid starbursts \citep{gla99,sul01}.  However,
\citet{bel01} find this explanation insufficient, and argue for an increasing
attenuation of the UV emission relative to H$\alpha$ for higher luminosity
galaxies.

\subsection{SFR$_{\mbox{\rm [O~{\sc ii}]}}$ and SFR$_{{\rm H}\alpha}$}

Although it is well known that the observed (i.e.\ uncorrected for extinction)
[O~{\sc ii}]/H$\alpha$ ratio shows a large scatter in local star-forming
galaxies (a factor of $\sim 7$), this scatter improves when extinction
corrected values are used \citep{jan01,ara02}. In our case
(Fig.~\ref{figure_sfrs}b) the comparison between both SFR estimates is, at
first sight, quite good within error bars. However, it is interesting to note
that the galaxy that deviates the most from the 1:1 relation is precisely the
object with lowest metallicity, namely GSS084\_4515. That metallicity is an
underlying cause of scatter when comparing [O~{\sc ii}]$\lambda3727$ with
H$\alpha$ luminosities has been suggested by \citet{jan01}.  To explore this
issue in greater detail, we compare in Fig.~\ref{figure_jansen} the correlation
found by these authors between the [O~{\sc ii}]/H$\alpha$ ratio and the
metallicity parameter R$_{23}$ for local emission-line galaxies which are not
dominated by an active galactic nucleus. The solid line is the linear fit to
the local galaxy sample, which is given by \citep{jan01}
\begin{equation}
\log(\mbox{[O~{\sc ii}]}/{\rm H}\alpha) = 0.82 \log {\rm R}_{23} - 0.48.
\end{equation}
The galaxies in our sample follow very well this trend. After applying a
metallicity correction to the [O~{\sc ii}] SFR estimates
(Fig.~\ref{figure_oii_ha_sfr}) we find an excellent match between both SFR
indicators.

\subsection{SFR$_{\rm IR}$, SFR$_{\rm radio}$ and SFR$_{{\rm H}\alpha}$}
\label{comparison_ir_halpha}

\begin{figure}
\epsscale{0.80}
\plotone{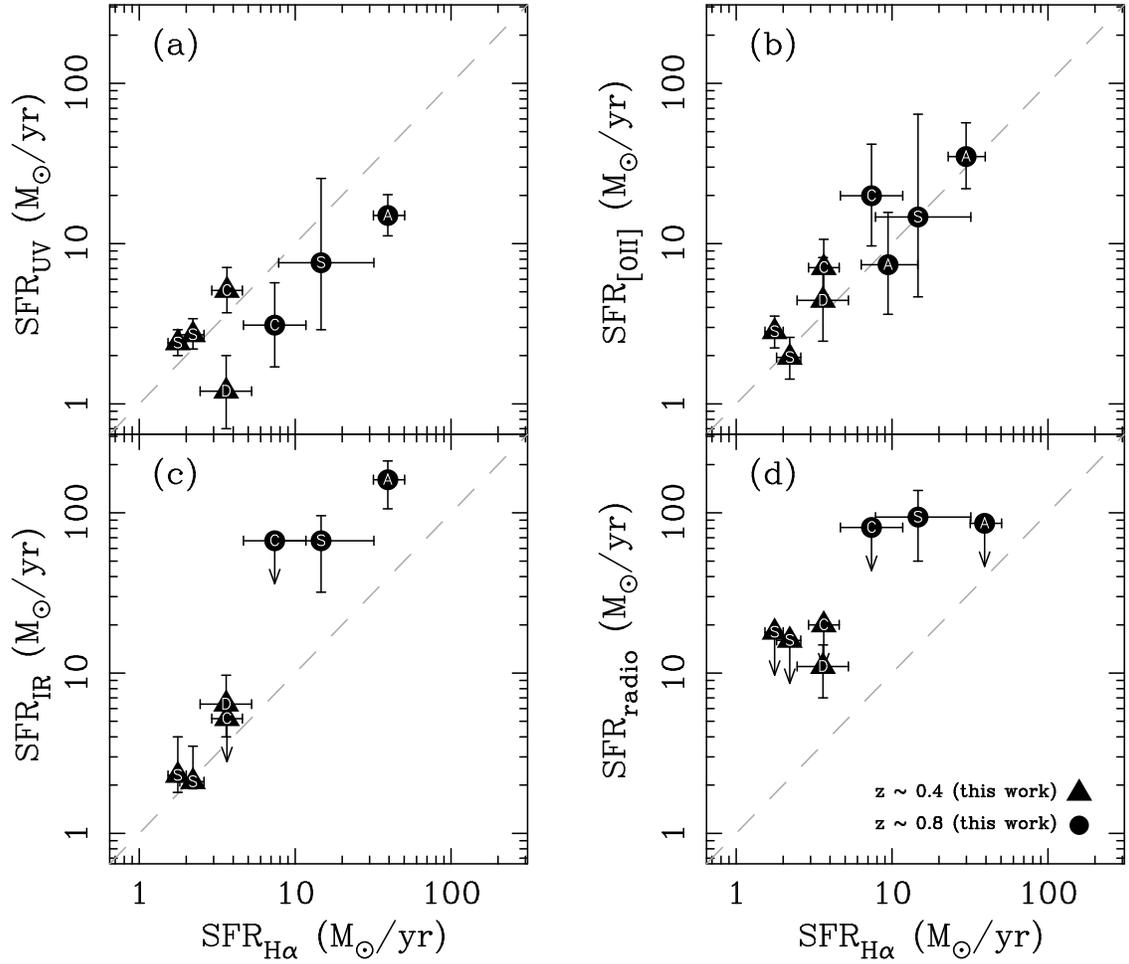}
\figcaption{Comparison of the five SFR indicators listed in
Table~\ref{table_sfrs}. Triangles correspond to the HDF-N galaxies ($z \sim
0.4$) and circles to the GSS subsample ($z \sim 0.8$). Symbols are labeled as
in Fig.~\ref{plot_el}. Upper limits in SFR$_{\rm IR}$ and SFR$_{\rm radio}$ are
shown with arrows. See discussion in
Sect.~\ref{sfr_comparison}\label{figure_sfrs}}
\end{figure}

\begin{figure}
\epsscale{0.85}
\plotone{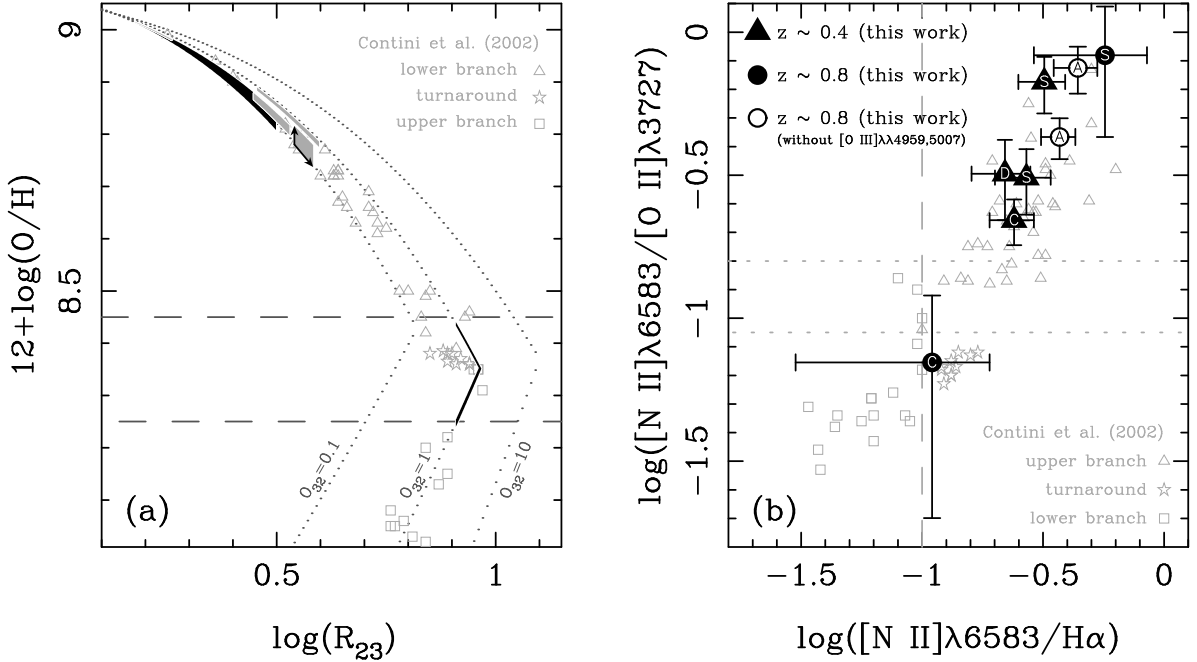}
\figcaption{{\bf Panel~(a)}: Calibration of oxygen abundance, as a function of
the line ratio R$_{23}\;\equiv\;$ ([O~{\sc ii}]$\lambda$3727+[O~{\sc
iii}]$\lambda\lambda$4959,5007)/H$\beta$, and parameterized by the ionization
parameter O$_{32}\;\equiv\;$ ([O~{\sc iii}]$\lambda\lambda$4959,5007)/[O~{\sc
ii}]$\lambda$3727. The dotted lines correspond to the analytic fits by
\citet{kob99} to the \citet{mcg91} set of photoionization models. The shaded
areas are the galaxies of our sample (lighter regions galaxies at $z \sim 0.4$;
for hd2-264.2 we only have lower limits in R$_{23}$ and O$_{23}$, which have
opposite effects in metallicity, as shown by the arrows).  The dashed lines
mark the turnaround region, where the uncertainties are large. Data points
come from Fig.~4 of \citet{con02}: open squares and triangles discriminate
between the lower and upper branches, whereas the open stars are objects that
fall in the intermediate region.  {\bf Panel~(b)}: Diagnostic diagram employed
to break the degeneracy in panel~(a).  We have also included in this diagram
the galaxy sample of \citet{con02}, which illustrates the method to break the
degeneracy of the oxygen abundance as a function of the R$_{23}$ index:
triangles correspond to their galaxies that fall on the upper branch of the
R$_{23}$ calibration ($\log(\mbox{[N~{\sc ii}]}\lambda6583/{\rm H}\alpha) >
-1$, and $\log(\mbox{[N~{\sc ii}]}\lambda6583/\mbox{[O~{\sc ii}]}\lambda3727) >
-1.05$), squares to objects that lie on the lower branch ($\log(\mbox{[N~{\sc
ii}]}\lambda6583/{\rm H}\alpha) < -1$, and $\log(\mbox{[N~{\sc
ii}]}\lambda6583/\mbox{[O~{\sc ii}]}\lambda3727) < -0.8$), whereas stars
correspond to galaxies that fall in the turnaround region of the calibration.
As is clear from this figure, the metallicity for the galaxies of our sample
must be read from the upper branch calibration of panel~(a), exception made for
GSS084\_4515, which lies in the intermediate region.
\label{figure_metallicities}}
\end{figure}

\begin{figure}
\epsscale{0.60}
\plotone{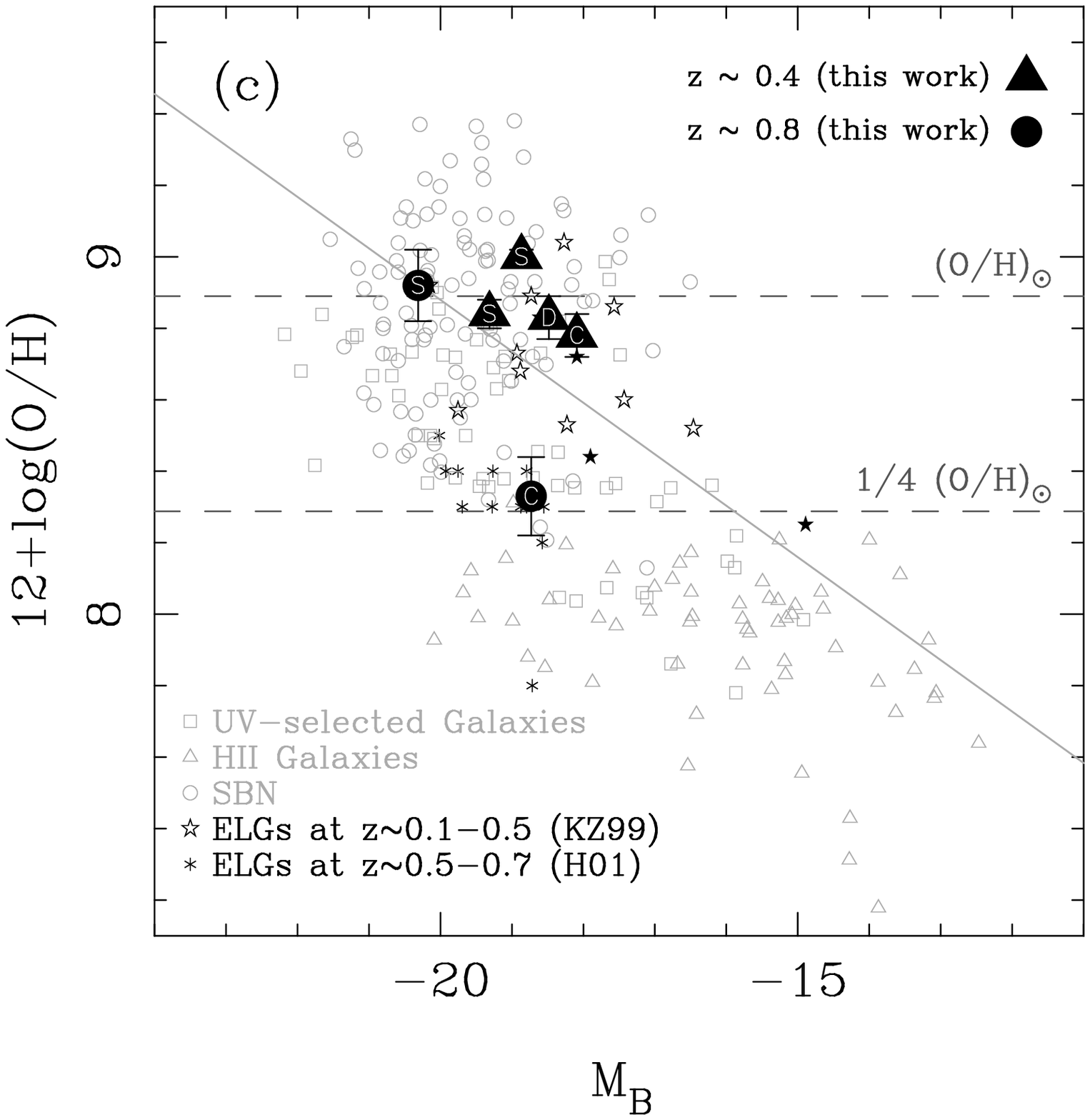}
\figcaption{Metallicity--luminosity diagram. The different galaxy samples have
been extracted from \citet[][Fig.~10a]{con02}. Note that in order to perform
the comparison, the rest-frame absolute $B$ magnitudes of the galaxies of our
paper have been transformed to a cosmology with $H_0 = 100 \; {\rm km} \; {\rm
s}^{-1} \; {\rm Mpc}^{-1}$, and $q_0=0.5$. The comparison samples include
non-local emission line galaxies from \citet[][4 CNELGs of their sample are
represented with filled stars]{koz99} and 14 luminous compact galaxies from
\citet{ham01}. The solid line is the least-squares fit to local irregular and
spiral galaxies from \citet[][Fig.~4]{koz99}.\label{figure_metlumi}}
\end{figure}

\begin{figure}
\epsscale{0.90}
\plotone{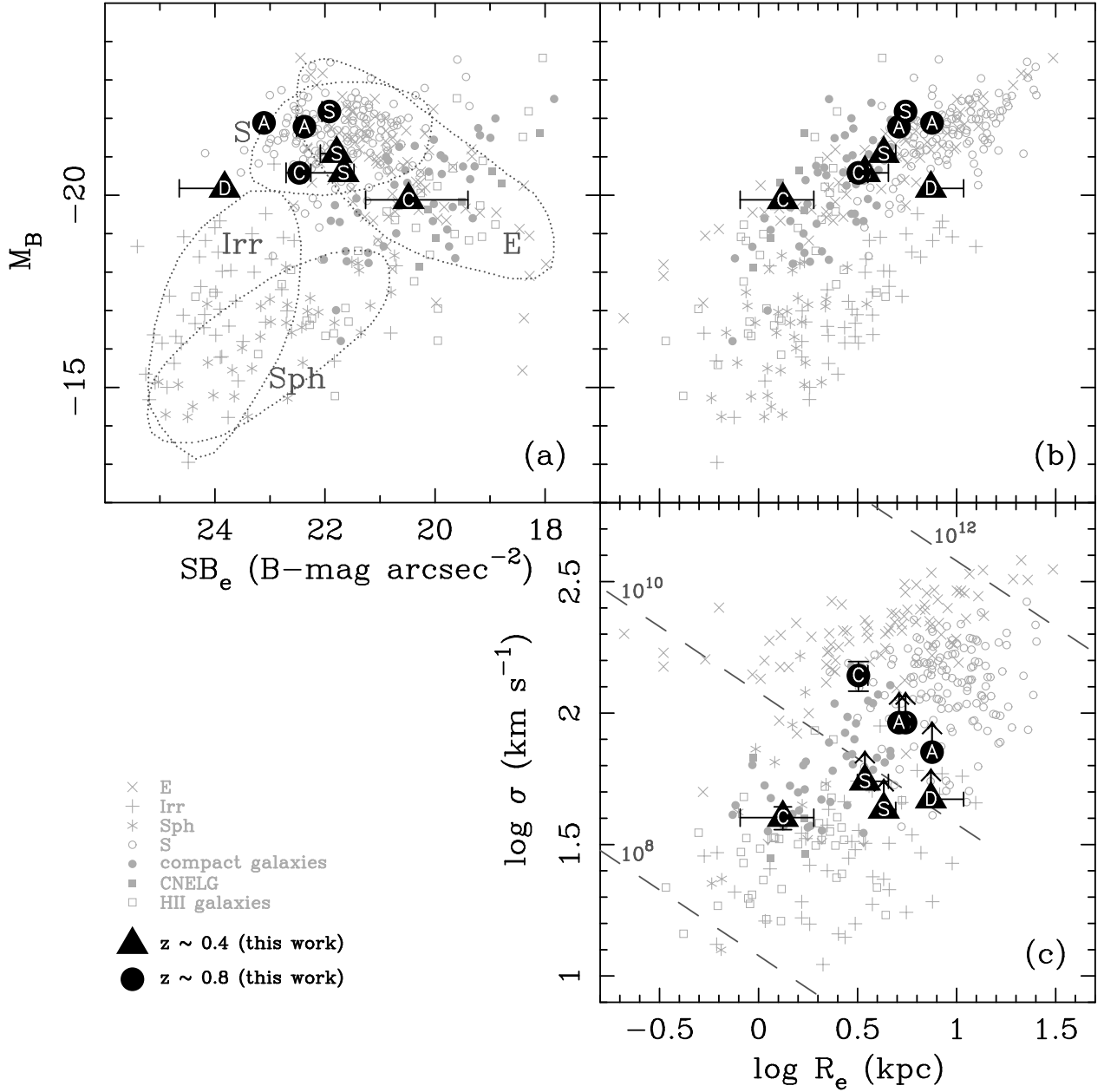}
\figcaption{Comparison between blue luminosities, half-light surface
brightnesses and velocity widths of our galaxy sample and the data from
\citet{phi97}. {\bf Panel~(a)}: Rest-frame surface brightness (averaged within
the half-light radius) and blue luminosities. The dotted regions indicate the
plot domain spanned by different types of local galaxies.  {\bf Panel~(b)}:
Comparison of rest-frame half-light radii and blue luminosities.  {\bf
Panel~(c)}: Comparison of rest-frame half-light radii and velocity widths.
Since the clear spiral-type galaxies of our sample are viewed almost face on,
we assume that the measured emission-line widths are underestimating the actual
rotational velocity of these objects. For that reason they are plotted as lower
limits.  The dashed lines are iso-mass tracks (in solar units) corresponding to
the virial mass estimation used by \citet{guz96}, $M \simeq 3 c_2/G\sigma^2
R_{\rm e}$, where we have assumed the exponential case $c_2=1.6$ for the
geometry dependent parameter. In this comparison we transformed our data to a
cosmology with $H_0 = 50 \; {\rm km} \; {\rm s}^{-1} \; {\rm Mpc}^{-1}$, and
$q_0=0.05$.
\label{figure_global}}
\end{figure}

\begin{figure}
\epsscale{0.60}
\plotone{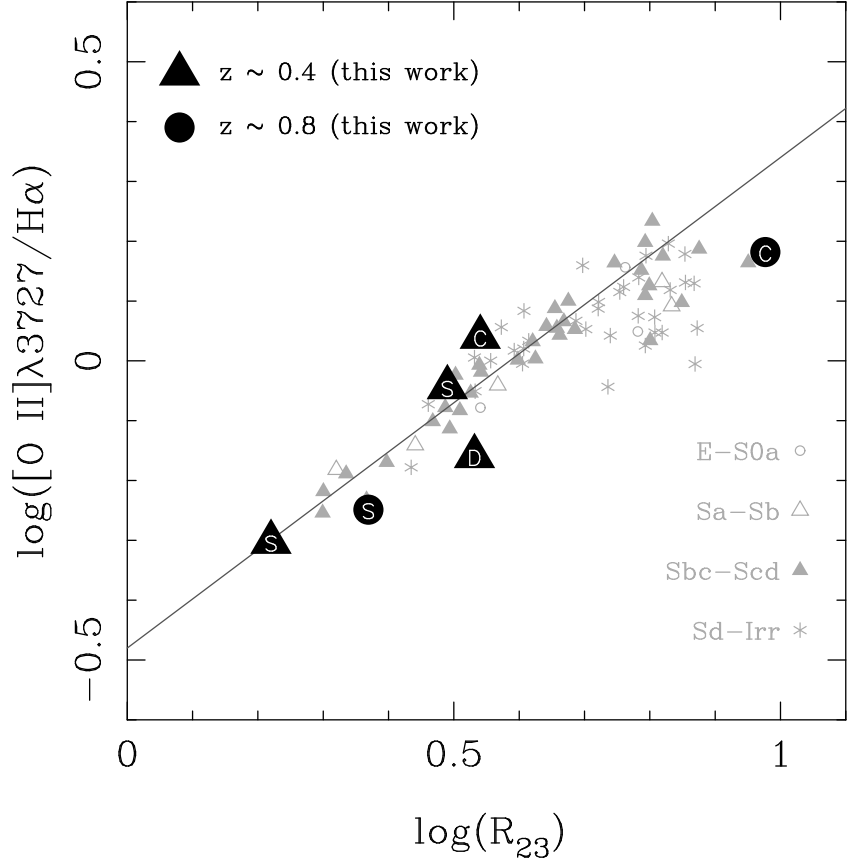}
\figcaption{Reddening-corrected [O~{\sc ii}]$\lambda3727$/H$\alpha$ ratio as a
function of the metallicity parameter R$_{23}$. Small symbols correspond to
local galaxies from \citet{jan01}, and the full line is a linear fit to these
data. The galaxies of our sample do follow very well the same relation.
\label{figure_jansen}}
\end{figure}

\begin{figure}
\epsscale{0.70}
\plotone{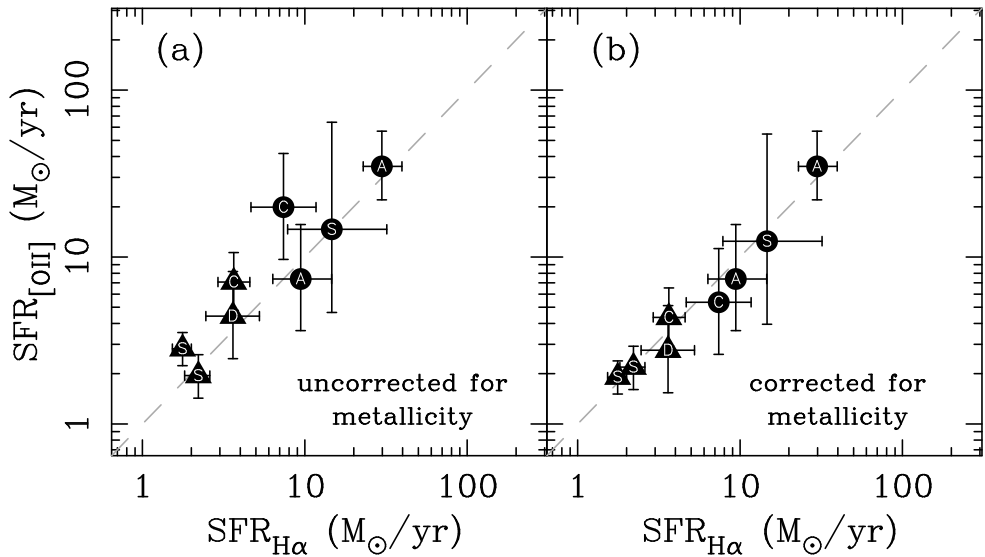}
\figcaption{Comparison between the SFRs derived from [O~{\sc ii}] and H$\alpha$
luminosities without any further correction ---panel~(a)--- and employing the
metallicity correction $\log(\mbox{[O~{\sc ii}]}/{\rm H}\alpha) = 0.82 \log{\rm
R}_{23}-0.48$ \citep{jan01} ---panel~(b)--- shown in Fig.~\ref{figure_jansen}.
Note that although for the antenae-like object (the pair of galaxies
GSS073\_1810) we have applied no correction since R$_{23}$ is unknown, the
locus of this system in Fig.~\ref{figure_metallicities}b suggests that its
R$_{23}$ parameter should be close to that of GSS084\_4515, hd4-656.1 and
hd4-795.111, for which the metallicity corrections in this diagram are small.
\label{figure_oii_ha_sfr}}

\end{figure}

Finally, in Fig.~\ref{figure_sfrs}c and~\ref{figure_sfrs}d we compare the
unbiased SFR$_{\rm IR}$ and SFR$_{\rm radio}$ estimators, with SFR$_{{\rm
H}\alpha}$. For the two galaxies of our sample with radio detections, SFR$_{\rm
IR}$ and SFR$_{\rm radio}$ agree very well within the error bars. The remaining
upper limits in radio fluxes also give consistent results with the IR data,
excluding GSS073\_1810, for which the IR flux is unexpectedly higher than the
constraint imposed by the radio flux. For the latter galaxy, an inspection of
the radio map of \citet[][Fig.~2, NW quadrant]{fom91} reveals that there is
indeed a detection (displayed as an isocontour at a level of 8$\mu$~Jy) at the
expected location of this object. The fact that it constitutes an interacting
system with a physical star-forming extension which is likely larger than that
exhibited by a single galaxy like GSS084\_4521, might explain why GSS073\_1810
does not present a concentrated peak flux density above the 16$\mu$~Jy
threshold used by \citet{fom91} to generate their ``complete'' catalog. Thus,
the integrated flux density for this object is probably very close to the
sensitivity limit of their survey. This result, together with the uncertainty
in the total infrared luminosity, lessens the discrepancy between SFR$_{\rm
IR}$ and SFR$_{\rm radio}$ for this object.

The most intriguing result, which is specially manifest in
Fig.~\ref{figure_sfrs}c, is the good agreement between these SFR indicators for
the galaxy subsample at $z \sim 0.4$, while for the higher redshift (and more
luminous) objects the agreement is very poor. To investigate this paradoxical
behavior in greater detail we compare in Fig.~\ref{figure_psfr} the ratio
SFR$_{\rm IR}$/SFR$_{{\rm H}\alpha}$ as a function of the IR luminosity. In
this diagram we have included additional galaxy samples, both local and at high
redshift. The small open symbols correspond to local star-forming galaxies from
\citet{gal96} and \citet[][excluding cluster galaxies and objects with apparent
diameters larger than 1.5~arcmin ---likely affected by aperture
effects---]{bua02}. The small solid squares represent the measurements of
galaxies observed by \citet{rig00} in the Hubble Deep Field South without a
correction for internal extinction. By including the average extinction
correction of these authors (a factor of 4 in the H$\alpha$ flux), their
measurements are displaced to the positions indicated by the tip of the arrows.
The H$\alpha$ fluxes in the latter sample were also not corrected for aperture
effects. We have ignored two galaxies from the Rigopoulou sample, namely
ISOHDF-S\_38 and ISOHDF-S\_39, because they are probable AGNs (both exhibit 
very strong [N~{\sc ii}] lines compared with H$\alpha$ ---see their Fig.~1---).
Asterisks are the extinction corrected objects from Fig.~3a of \citet{sul01},
where we have determined $L_{\rm IR}$ from their 1.4~GHz luminosities using the
relations shown in Sect.~\ref{sfr_from_radio} (we have excluded upper limits in
Sullivan et al.\ data). The dashed line is a bisector least-squares fit
\citep{iso90} to our galaxy sample,
\begin{equation} 
{\rm SFR}_{\rm IR}/{\rm SFR}_{{\rm H}\alpha} \simeq 
   3.3 \times 10^{-4} \; (L_{\rm IR}/L_\odot)^{0.35},
\label{sfrratio_lir} 
\end{equation} 
where we have excluded the upper limits in $L_{\rm IR}$ (and thus in SFR$_{\rm
IR}$/SFR$_{{\rm H}\alpha}$).  Our lower redshift ($z \sim 0.4$) galaxies are
compatible with the local reference sample, in the sense that SFR$_{{\rm
H}\alpha}$ traces almost all the SFR (if not all) as derived from the IR
luminosity. However, this is not true when the more luminous galaxies are
considered.  In particular, only one object, out of 10 local galaxies with
$L_{\rm IR} > 10^{11}$~L$_\odot$, falls above the relation given in
Eq.~(\ref{sfrratio_lir}). On the contrary, and although the actual extinctions
for the high-redshift sample of \citet{rig00} are unknown, the averaged
correction estimated by these authors seems to accommodate their galaxies in
the previous fit. Finally, the two most luminous IR objects from the sample of
\citet{sul01} do also appear quite close to the same fit. Summarizing, the data
we have collected here indicate that the extinction corrected SFR$_{{\rm
H}\alpha}$ estimates in $z \sim 0.8$ galaxies miss an increasing fraction of
the total SFR, which is clear evidence of the presence of highly obscured
dust-enshrouded star forming regions within these galaxies.  In addition, there
is an indication that this underestimate may be a function of $L_{\rm IR}$
(Eq.~\ref{sfrratio_lir}). However, considering the luminosity and redshift
segregation of our galaxy sample, we cannot yet preclude an effect due only to
$z$.

Interestingly, the two galaxies with compact aspect, hd2-264.2 and
GSS084\_4515, were both undetected by ISO, and thus we have only plotted upper
limits for these objects in Fig.~\ref{figure_psfr}. However, it is important to
highlight that this common morphology is misleading, since, as we have
previously discussed, these galaxies exhibit different metallicity, mass and
structural properties, apart from a difference of a factor of $\sim 10$ in
total far-infrared luminosity. The arrow in hd2-264.2 places this object close
to SFR$_{\rm IR}$/H$\alpha \simeq 1$, whereas in the case of the LCG
GSS084\_4515 its real $L_{\rm IR}$ would put it either on the fit given by
Eq.~(\ref{sfrratio_lir}), or in the locus of the $z \sim 0.4$ objects. If, as
suggested by \citet{ham01}, LCGs are evolved starbursts and the progenitors of
the present-day spiral bulges, objects of this type may have an important
contribution to the SFR density at intermediate redshifts. These authors have
also predicted the total far-infrared luminosities of LCGs to be around
$10^{11}$~L$_\odot$, near or slightly below the limiting sensitivity reached by
ISO in the redshift range $0.5 \la z \la 0.7$. If we assume that GSS084\_4515
belongs to the same galaxy family, and although its non-detection by ISO is
then not surprising, this galaxy will accommodate in the fit of
Eq.~(\ref{sfrratio_lir}). This result highlights the relevance of quantifying
the actual $L_{\rm IR}$ of LCGs, which will be possible in a near future with
the help of SIRTF.

Another important question is whether it is possible to use any measurable
parameter exclusively derived from the optical lines with the aim of correcting
the biased SFR$_{{\rm H}\alpha}$ estimates, in order to match the values
obtained from the total far-infrared luminosity. In this sense, and as we
already mentioned in Sect.~\ref{introduction}, \citet{hop01} and \citet{sul01}
have recently reported a correlation between the optical extinction (derived
from the Balmer decrement) and the star formation rate (or luminosity) of the
galaxies. \citet{bua02} also find correlations between the measured
extinction in H$\alpha$, A[H$\alpha$], and $L_{\rm IR}$, although with a very
large scatter.

We have represented in Fig.~\ref{figure_sfr_amag} the ratio SFR$_{\rm
IR}$/SFR$_{{\rm H}\alpha}$ as a function of A[H$\alpha$].  For local galaxies
this ratio correlates with A[H$\alpha$] when H$\alpha$ fluxes uncorrected for
extinction are used.  After correcting the H$\alpha$ flux for extinction, local
galaxies and our galaxy subsample at $z \sim 0.4$ scatter around SFR$_{\rm
IR}$/SFR$_{{\rm H}\alpha} \sim 1$, whereas the $z \sim 0.8$ objects remain
above this value.  Since the measured H$\alpha$ extinctions in
GSS073\_1810 and in GSS084\_4521 are very similar, we cannot use this parameter
to conclude whether there is any correlation between the systematic
underestimation of SFR from H$\alpha$ as a function of A[H$\alpha$].  Anyway,
we should note that these two galaxies have H$\alpha$ extinctions similar to
hd2-264.1, and lower than hd2-264.2. Both hd2-264.1 and hd2-264.2 have total
far-infrared luminosities one order of magnitude fainter that either
GSS073\_1810 and GSS084\_4515. For that reason, if there were actually a
correlation between the observed rest-frame optical extinction and the ratio
SFR$_{\rm IR}$/H$\alpha$, it should be sought in galaxies with $L_{\rm IR} >
10^{11}$~L$_\odot$ and, presumably, at intermediate redshifts. Clearly more
data are required in order to settle this question.

\section{Conclusions}
\label{conclusions}

In this work we have compared several star formation rate estimators using a
small but diverse sample of galaxies at two intermediate redshifts
($z \sim 0.4$ and $z \sim 0.8$).
By selecting the sample galaxies on the basis
of their ISOCAM-15~$\mu$m mid-infrared luminosities, we have sampled SFRs
ranging from 2 to 160~M$_\odot {\rm yr}^{-1}$. Their morphological types
include spiral, compact and merging emission-line galaxies. It is important to
highlight that the redshift segregation of the sample is also accompanied by
a segregation in total far-infrared luminosity, with the farthest galaxies 
being the most luminous. This fact leads to an ambiguity between luminosity and
distance that must be kept in mind when interpreting the results.

The observed spectral range includes the most prominent
optical emission lines, from [O~{\sc ii}]$\lambda$3727 to [S~{\sc
ii}]$\lambda\lambda$6716,6731. The availability of H$\beta$ and H$\alpha$
allows the determination of color excesses in the nebular gas and
correcting the emission-line fluxes for extinction. The diagnostic diagrams
generated from the corrected emission-line ratios confirm the star-forming
nature of the galaxies, as it was already suggested by evidence from
their total far-infrared luminosity, observations in the X-ray domain, and the
absence of broad emission lines. In addition, the overall characteristics of
the selected galaxies, in particular their metallicity and structural
properties, also match those of star forming galaxies, both local and at
intermediate redshifts. 

The good agreement between HST and ground-based photometry, in one hand, and
the spectrophotometric ca\-li\-bra\-tion of the spectra, on the other, allows
us to be confident on the reliability of the estimated aperture corrections,
the final optical emission-line luminosities, and thus the SFRs from them
obtained. Our results have shown the following.

(i) There is a general good agreement in the comparison between the five
available SFR indicators (from UV, [O~{\sc ii}], H$\alpha$, IR and radio
luminosities) for the $z \sim 0.4$ (L$_{\rm IR} < 10^{11}$~L$_\odot$) galaxies,
whereas the situation is not the same in the case of the $z \sim 0.8$ (L$_{\rm
IR} > 10^{11}$~L$_\odot$) subsample, being the discrepancies different
depending on the considered SFR estimator. Focusing in the $z \sim 0.8$
galaxies, the SFRs derived from UV, [O~{\sc ii}] and IR luminosities are,
respectively, lower, similar and higher, than the values obtained from
H$\alpha$.  The paradoxical behavior of SFR$_{\rm UV}$ versus SFR$_{{\rm
H}\alpha}$ is still subject of debate, and among the possible explanations for
the observed discrepancies are temporal variations in the star formation
history and luminosity dependent attenuations of the UV emission.  From the
observational perspective, we confirm the findings of \citet{sul01} and extend
their result up to $z \sim 0.8$.

(ii) The correction for metallicity of the [O~{\sc ii}]/H$\alpha$ ratio (using
the relation found in local galaxies) greatly improves the concordance between
the SFRs derived from the extinction corrected luminosities of both emission
lines.  The correction does in fact work for the galaxies in our sample in the
two redshift (or luminosity) bins.

\begin{figure}
\epsscale{0.80}
\plotone{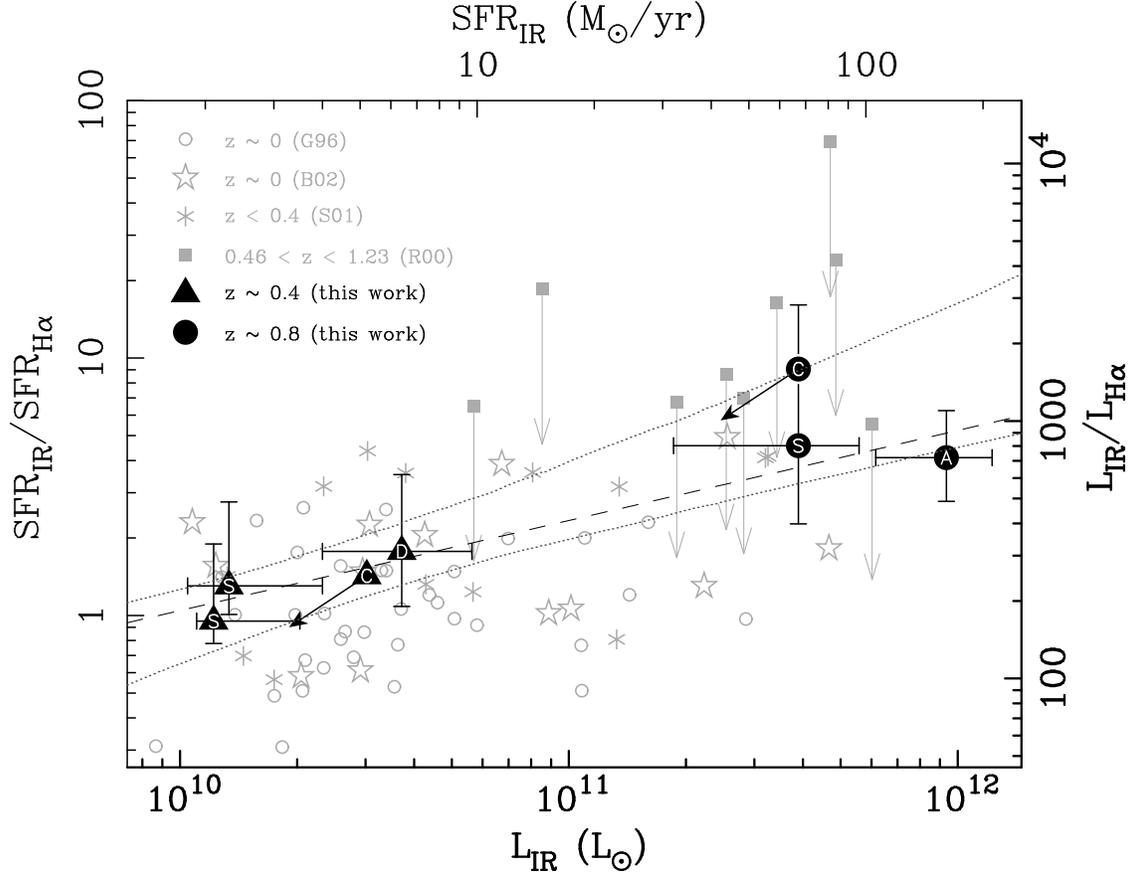}
\figcaption{Ratio of SFR$_{\rm IR}$ over extinction-corrected SFR$_{{\rm
H}\alpha}$, as a function of $L_{\rm IR}$ ---the horizontal scale at the top
gives the SFR$_{\rm IR}$ as computed from Eq.~(\ref{equation_sfr_ir})---.
Large filled symbols are the galaxies of our sample (triangles for objects with
$z\sim 0.4$, and circles for galaxies with $z\sim 0.8$; symbols labeled as in
Fig.~\ref{plot_el}), with SFR$_{{\rm H}\alpha}$ corrected for extinction and
aperture as explained in the text.  Small filled squares are the galaxies
observed by \citet{rig00} in the Hubble Deep Field South: the data points
correspond to the values without extinction correction, while the tips of the
arrows indicate the effect of using the average extinction correction (a factor
of 4 in the H$\alpha$ flux) employed by Rigopoulou et al.\ in their work; the
H$\alpha$ fluxes in this last sample were not corrected either for aperture
effects.  Asterisks are the extinction corrected objects from Fig.~3a of
\citet{sul01}, whereas open circles and stars correspond to local galaxies from
\citet{gal96,gal97} and \citet[][excluding cluster galaxies and objects with
apparent diameter larger than 1.5~arcmin]{bua02}, respectively. The dashed line
is a bisector least-squares fit to our galaxy sample, excluding upper limits in
$L_{\rm IR}$ (in both axis), whereas the dotted lines indicate the 1~$\sigma$
error in the fit prediction derived from numerical simulations via error
bootstrapping.  \label{figure_psfr}}
\end{figure}

\begin{figure}
\epsscale{1.00}
\plotone{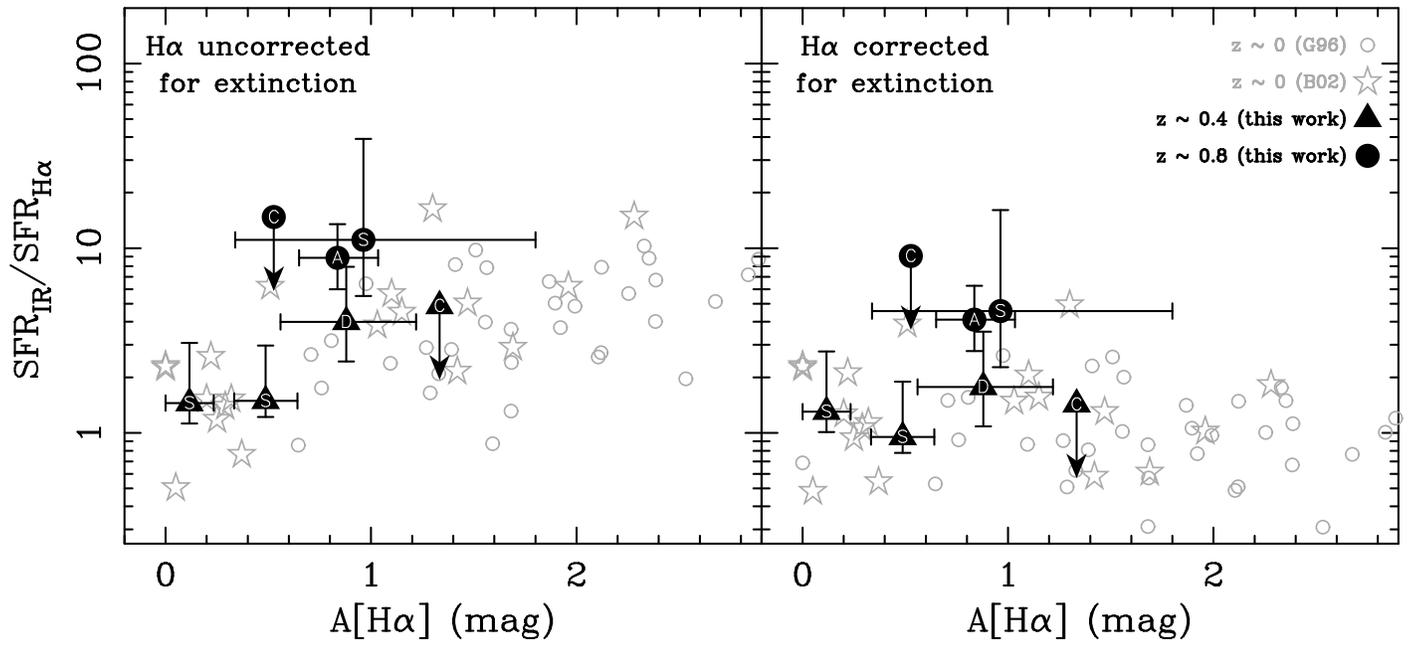} \figcaption{Ratio of SFR$_{\rm IR}$ over SFR$_{{\rm
H}\alpha}$ (with H$\alpha$ not corrected for extinction in the left panel, and
corrected in the right panel) as a function of the measured extinction in
H$\alpha$. Symbols as in Fig.~\ref{figure_psfr}.\label{figure_sfr_amag}}
\end{figure}

(iii) The fit given in Eq.~(\ref{sfrratio_lir}), and shown in
Fig.~\ref{figure_psfr}, indicates that {\it extinction corrected\/} SFR$_{{\rm
H}\alpha}$ estimates in luminous infrared galaxies at intermediate redshifts
miss an increasing fraction of the total SFR, and that the degree of
underestimation increases with $L_{\rm IR}$. This result confirms the finding
by \citet{rig00} who used an averaged extinction correction of $\sim 4$ derived
from $V-K$ colors. In this work we quantify more rigorously this effect by
employing line fluxes corrected for extinction making use of color excesses
computed from the Balmer decrement.  Thus, there is here an observational
evidence that the Balmer emission lines, H$\alpha$ and H$\beta$, do not probe
the same region of a galaxy than the one which is responsible for the strong IR
luminosity in the case of luminous IR galaxies. This is even more remarquable
in Fig.~\ref{figure_sfr_amag} where galaxies with very different SFRs share the
same A[H$\alpha$]. The spatial resolution in the MIR is not sufficient to
confirm this statement at these redshifts, but the case of the Antennae galaxy
NGC~4038/4039 \citep{mir98} offers a perfect example of such behavior in the
local universe. A possible explanation for this effect is that a fraction of
the star formation in these galaxies is embedded in dense and opaque dust
clouds and that this fraction increases with the total star formation.  As
already expected, galaxy encounters (hd2-264.1 and hd2-264.2) and mergers
(GSS073\_1810) are the likely causes for triggering star formation episodes.
Whether a quantitative estimation of the hidden star formation is definitely
inaccessible to rest-frame optical spectroscopic studies, and in particular to
the use of the measured H$\alpha$ extinction, seems unclear, and deserves
further research work.

Some kind of evolution in galaxies with $L_{\rm IR} > 10^{11}$~L$_\odot$ at $z
\la 0.4$ cannot be discarded, since, as we have shown in
Fig.~\ref{figure_psfr}, an important fraction of local galaxies with those
luminosities does not show such a large underestimation of the SFR when using
H$\alpha$ fluxes instead of far-infrared luminosities. However the scarcity of
the galaxy samples strongly demands additional observational work in order to
settle this point.

The inclusion of two apparently compact galaxies in our sample has allowed us
to analyze the relation between the different SFR estimators applied to this
kind of object relative to what is measured for
spiral-like galaxies. In particular, one of these two objects
qualifies as a luminous compact galaxy, with similar properties to those
of the LCGs at intermediate redshifts studied by \citet{ham01}. Surprisingly,
this object may be suffering similar dust-enshrouded star formation processes
as galaxies with a very different morphological aspect, like the bona fide
spiral GSS084\_4521 or even the antenae-like colliding system GSS073\_1810. The
confirmation of this still speculative result, which has important implications
concerning the accurate estimation of the cosmic SFR density at intermediate
redshifts, will be possible after the launch of SIRTF.

\acknowledgments

Valuable discussions with Jaime Zamorano, Pablo P\'{e}rez-Gonz\'{a}lez and
Ranga-Ram Chary are gratefully acknowledged. We thank Herv\'e Aussel for
facilitating us the deconvolved ISCOCAM flux for hd2-264.1 and Hector Flores
for providing us with his revised ISOCAM fluxes for the GSS galaxies. We are
grateful to the staff of the W.~M.~Keck Observatory for their help during the
observations. This research has made use of the SIMBAD database, operated at
CDS, Strasbourg, France.  Support for this work has been provided by NSF grants
AST~95-29028 and AST~00-71198.  N.C.\ acknowledges financial support from a UCM
Fundaci\'{o}n del Amo fellowship, a short contract at UCSC, and from the
Spanish Programa Nacional de Astronom\'{\i}a y Astrof\'{\i}sica under grant
AYA2000-977. D.E.\ wishes to thank the American Astronomical Society for its
support through the Chretien International Research Grant and Joel Primack and
David Koo for supporting his research through NASA grants NAG5-8218 and
NAG5-3507. R.P.S.\ acknowledges support provided by the National Science
Foundation through grant GF-1002-99 and from the Association of Universities
for Research in Astronomy, Inc., under NSF cooperative agreement AST~96-13615,
and CNPq/Brazil, for financial support through grant 200510/99-1.  J.G.\
acknowledges financial support from the Spanish Programa Nacional de
Astronom\'{\i}a y Astrof\'{\i}sica under grant AYA2000-1790.



\begin{deluxetable}{lcccccc}
\tablecaption{Galaxy sample.\label{table_sample}}
\tablewidth{0pt}
\tablehead{
\colhead{Galaxy\tablenotemark{a}} & 
\colhead{R.A.\tablenotemark{b}} &
\colhead{DEC.\tablenotemark{b}} &
\colhead{$z_{\rm spec}$\tablenotemark{c}} &
\colhead{E$_{\rm G}$(B$-$V)\tablenotemark{d}} &
\colhead{Run\tablenotemark{e}} &
\colhead{t$_{\rm exp}$\tablenotemark{f}}
}
\startdata
hd2-264.1     & 
12 36 49.76 & 62 13 13.1 & 
0.475 & 
0.012 &  
4 (0.75\arcsec)  & 1200 \\
hd2-264.2     & 
12 36 49.38 & 62 13 11.2 & 
0.477 & 
0.012 &  
4 (0.75\arcsec)  & 1200 \\
hd4-656.1     & 
12 36 42.91 & 62 12 16.3 & 
0.454 & 
0.012 &  
4 (1.25\arcsec)  &  900 \\
hd4-795.111   & 
12 36 41.95 & 62 12 05.4 & 
0.433 & 
0.012 &  
4 (1.25\arcsec)  &  900 \\
GSS073\_1810  & 
14 17 42.64 & 52 28 45.3 & 
0.831 & 
0.007 & 
1 (1.18\arcsec), 5 (0.76\arcsec) & 3000, 2400 \\
GSS084\_4515  & 
14 17 40.43 & 52 27 19.4 & 
0.812 & 
0.008 & 
3 (1.00\arcsec), 5 (0.76\arcsec) & 3600, 2400 \\
GSS084\_4521  & 
14 17 40.55 & 52 27 13.6 & 
0.754 & 
0.008 & 
2 (1.00\arcsec), 5 (0.76\arcsec) & 3000, 2400 
\enddata

\tablenotetext{a}{Galaxy identification: for the HDF galaxies we follow
\citet{wil96}, whereas for the GSS galaxies we employ a name based on the field
and the WFPC chip numbers, and the and first two digits of the X and Y centroid
coordinates.}

\tablenotetext{b}{J2000.0 coordinates for the HDF (drizzled Version~2) and GSS
\citep{rho94} galaxies.}

\tablenotetext{c}{Spectroscopic redshifts obtained from measurements of 
individual spectral features, and/or cross-correlation technique.}

\tablenotetext{d}{Galactic color excesses derived from the dust maps of
\citet{sch98} and the software available at
\url{http://astron.berkeley.edu/davis/dust/}.}

\tablenotetext{e}{Spectroscopic observing runs with Keck: 1=LRIS (April~1996),
2=LRIS (May~1997), 3=LRIS (April~1998), 4=ESI (May~2000) and 5=NIRSPEC 
(July~2000). Slit widths employed in each case are given within parenthesis.}

\tablenotetext{f}{Total exposure times (seconds).}

\end{deluxetable}


\begin{deluxetable}{l@{}c@{}c@{}c@{}c@{}c@{}c@{}c@{}c@{}c@{}c@{}}
\tabletypesize{\footnotesize}
\rotate
\tablecaption{Measured emission-line fluxes and color 
excesses.\label{table_fluxes}}
\tablewidth{0pt}
\tablehead{
\colhead{Galaxy} & 
\colhead{$F({\rm\mbox{[O~{\sc ii}]}})$} &
\colhead{$F({\rm H}\beta)$} &
\colhead{$F({\rm\mbox{[O~{\sc iii}]}})$} &
\colhead{$F({\rm\mbox{[O~{\sc iii}]}})$} &
\colhead{$F({\rm\mbox{[N~{\sc ii}]}})$} &
\colhead{$F({\rm H}\alpha)$} &
\colhead{$F({\rm\mbox{[N~{\sc ii}]}})$} &
\colhead{$F({\rm\mbox{[S~{\sc ii}]}})$} &
\colhead{$F({\rm\mbox{[S~{\sc ii}]}})$} &
\colhead{E(B$-$V)$_{\rm gas}$} \\
&
\colhead{\makebox[0in][c]{$\lambda\lambda$3727,3729}} &
\colhead{$\lambda$4861} &
\colhead{$\lambda$4959} &
\colhead{$\lambda$5007} &
\colhead{$\lambda$6549} &
\colhead{$\lambda$6563} &
\colhead{$\lambda$6583} &
\colhead{$\lambda$6717} &
\colhead{$\lambda$6734} &
\colhead{}
}
\startdata
$\begin{array}{@{}c}\mbox{hd2-264.1} \\ \,\end{array} $ &   
$\begin{array}{@{}c@{\;}c@{}} 4.49 & (2) \\ 0.55 \end{array} $ &  
$\begin{array}{@{}c@{\;}c@{}} 3.33 & (1) \\ 0.44 \end{array} $ &  
$\begin{array}{@{}c@{\;}c@{}} 1.11 & (1) \\ 0.72 \end{array} $ &  
$\begin{array}{@{}c@{\;}c@{}} 3.99 & (1) \\ 0.22 \end{array} $ &  
$\begin{array}{@{}c@{\;}c@{}} 2.46 & (1) \\ 2.13 \end{array} $ &  
$\begin{array}{@{}c@{\;}c@{}} 14.7 & (1) \\ 1.4 \end{array} $  &  
$\begin{array}{@{}c@{\;}c@{}} 3.23 & (1) \\ 0.72 \end{array} $ &  
$\begin{array}{@{}c@{\;}c@{}} 0.40 & (3) \\ 0.41 \end{array} $ &  
$\begin{array}{@{}c@{\;}c@{}} 1.62 & (3) \\ 0.98 \end{array} $ &  
$ 0.37^{+0.15}_{-0.14} $ \\                                 
$\begin{array}{@{}c}\mbox{hd2-264.2} \\ \,\end{array} $ &  
$\begin{array}{@{}c@{\;}c@{}} 4.98 & (2) \\ 0.25 \end{array} $ &  
$\begin{array}{@{}c@{\;}c@{}} 2.72 & (1) \\ 0.25 \end{array} $ &  
$\begin{array}{@{}c@{\;}c@{}} 0.21 & (3) \\ 0.23 \end{array} $ &  
$\begin{array}{@{}c@{\;}c@{}}
  \makebox[0in][r]{$>$}0.63 & (1) \\\mbox{---}\end{array} $ &  
$\begin{array}{@{}c@{\;}c@{}} 1.27 & (3) \\ 0.58 \end{array} $ &  
$\begin{array}{@{}c@{\;}c@{}} 15.0 & (1) \\ 0.7 \end{array} $  &  
$\begin{array}{@{}c@{\;}c@{}} 3.61 & (1) \\ 0.69 \end{array} $ &  
$\begin{array}{@{}c@{\;}c@{}}\mbox{---}& \\\mbox{---}\end{array} $ & 
$\begin{array}{@{}c@{\;}c@{}} 1.00 & (3) \\ 0.50 \end{array} $ &  
$ 0.56^{+0.10}_{-0.09} $ \\                                 
$\begin{array}{@{}c}\mbox{hd4-656.1} \\ \,\end{array} $ &  
$\begin{array}{@{}c@{\;}c@{}} 17.9 & (2) \\ 0.9 \end{array} $  &  
$\begin{array}{@{}c@{\;}c@{}} 7.46 & (1) \\ 0.30 \end{array} $ &  
$\begin{array}{@{}c@{\;}c@{}} 1.11 & (3) \\ 0.61 \end{array} $ &  
$\begin{array}{@{}c@{\;}c@{}} 3.11 & (3) \\ 1.57 \end{array} $ &  
$\begin{array}{@{}c@{\;}c@{}}\mbox{---}& \\\mbox{---}\end{array} $ & 
$\begin{array}{@{}c@{\;}c@{}} 22.7 & (1) \\ 0.9 \end{array} $  &  
$\begin{array}{@{}c@{\;}c@{}} 6.18 & (3) \\ 1.60 \end{array} $ &  
$\begin{array}{@{}c@{\;}c@{}} 4.60 & (3) \\ 1.19 \end{array} $ &  
$\begin{array}{@{}c@{\;}c@{}} 2.70 & (3) \\ 0.77 \end{array} $ &  
$ 0.04^{+0.05}_{-0.04} $ \\                                 
$\begin{array}{@{}c}\mbox{hd4-795.111} \\ \,\end{array} $ &  
$\begin{array}{@{}c@{\;}c@{}} 6.29 & (2) \\ 0.55 \end{array} $ &  
$\begin{array}{@{}c@{\;}c@{}} 5.58 & (1) \\ 0.29 \end{array} $ &  
$\begin{array}{@{}c@{\;}c@{}} 0.35 & (3) \\ 0.16 \end{array} $ &  
$\begin{array}{@{}c@{\;}c@{}} 1.21 & (1) \\ 0.35 \end{array} $ &  
$\begin{array}{@{}c@{\;}c@{}}\mbox{---}& \\\mbox{---}\end{array} $ & 
$\begin{array}{@{}c@{\;}c@{}} 20.4 & (1) \\ 1.1 \end{array} $  &  
$\begin{array}{@{}c@{\;}c@{}} 6.59 & (1) \\ 1.34 \end{array} $ &  
$\begin{array}{@{}c@{\;}c@{}} 1.98 & (3) \\ 0.53 \end{array} $ &  
$\begin{array}{@{}c@{\;}c@{}} 1.57 & (3) \\ 0.42 \end{array} $ &  
$ 0.20^{+0.07}_{-0.06} $ \\                                 
$\begin{array}{@{}c}\mbox{GSS073\_1810A}\!\! \\ \,\end{array} $ &  
$\begin{array}{@{}c@{\;}c@{}} 7.46 & (2) \\ 0.10 \end{array} $ &  
$\begin{array}{@{}c@{\;}c@{}} 4.61 & (1) \\ 0.52 \end{array} $ &  
$\begin{array}{@{}c@{\;}c@{}}\mbox{---}& \\\mbox{---}\end{array} $ & 
$\begin{array}{@{}c@{\;}c@{}}\mbox{---}& \\\mbox{---}\end{array} $ & 
$\begin{array}{@{}c@{\;}c@{}} 2.98 & (3) \\ 0.68 \end{array} $ &  
$\begin{array}{@{}c@{\;}c@{}} 22.3 & (1) \\ 1.3 \end{array} $  &  
$\begin{array}{@{}c@{\;}c@{}} 8.37 & (1) \\ 1.28 \end{array} $ &  
$\begin{array}{@{}c@{\;}c@{}} 2.83 & (3) \\ 0.48 \end{array} $ &  
$\begin{array}{@{}c@{\;}c@{}} 4.70 & (3) \\ 0.78 \end{array} $ &  
$ 0.45^{+0.12}_{-0.11} $ \\                                 
$\begin{array}{@{}c}\mbox{GSS073\_1810B}\!\! \\ \,\end{array} $ &  
$\begin{array}{@{}c@{\;}c@{}} 3.87 & (2) \\ 0.10 \end{array} $ &  
$\begin{array}{@{}c@{\;}c@{}} 2.92 & (1) \\ 0.48 \end{array} $ &  
$\begin{array}{@{}c@{\;}c@{}}\mbox{---}& \\\mbox{---}\end{array} $ & 
$\begin{array}{@{}c@{\;}c@{}}\mbox{---}& \\\mbox{---}\end{array} $ & 
$\begin{array}{@{}c@{\;}c@{}} 1.65 & (3) \\ 0.63 \end{array} $ &  
$\begin{array}{@{}c@{\;}c@{}} 11.2 & (1) \\ 1.1 \end{array} $  &  
$\begin{array}{@{}c@{\;}c@{}} 4.88 & (1) \\ 0.90 \end{array} $ &  
$\begin{array}{@{}c@{\;}c@{}} 1.73 & (3) \\ 0.38 \end{array} $ &  
$\begin{array}{@{}c@{\;}c@{}} 2.03 & (1) \\ 0.83 \end{array} $ &  
$ 0.24^{+0.18}_{-0.16} $ \\                                 
$\begin{array}{@{}c}\mbox{GSS084\_4515} \\ \,\end{array} $ &  
$\begin{array}{@{}c@{\;}c@{}} 16.4 & (2) \\ 0.2 \end{array} $  &  
$\begin{array}{@{}c@{\;}c@{}} 4.35 & (1) \\ 0.62 \end{array} $ &  
$\begin{array}{@{}c@{\;}c@{}} 6.18 & (1) \\ 0.76 \end{array} $ &  
$\begin{array}{@{}c@{\;}c@{}} 15.3 & (1) \\ 0.5  \end{array} $ &  
$\begin{array}{@{}c@{\;}c@{}} 4.20 & (3) \\ 1.75 \end{array} $ &  
$\begin{array}{@{}c@{\;}c@{}} 16.2 & (1) \\ 2.2 \end{array} $  &  
$\begin{array}{@{}c@{\;}c@{}} 1.83 & (3) \\ 1.20 \end{array} $ &  
$\begin{array}{@{}c@{\;}c@{}} 1.00 & (3) \\ 1.32 \end{array} $ &  
$\begin{array}{@{}c@{\;}c@{}} 1.20 & (3) \\ 1.55 \end{array} $ &  
$ 0.22^{+0.17}_{-0.17} $ \\                                 
$\begin{array}{@{}c}\mbox{GSS084\_4521} \\ \,\end{array} $ &  
$\begin{array}{@{}c@{\;}c@{}} 5.05 & (2) \\ 0.09 \end{array} $ &  
$\begin{array}{@{}c@{\;}c@{}} 3.82 & (3) \\ 1.24 \end{array} $ &  
$\begin{array}{@{}c@{\;}c@{}} 0.38 & (3) \\ 0.28 \end{array} $ &  
$\begin{array}{@{}c@{\;}c@{}}\mbox{---}& \\\mbox{---}\end{array} $ & 
$\begin{array}{@{}c@{\;}c@{}} 2.53 & (1) \\ 1.88 \end{array} $ &  
$\begin{array}{@{}c@{\;}c@{}} 17.6 & (1) \\ 1.9 \end{array} $  &  
$\begin{array}{@{}c@{\;}c@{}} 10.1 & (1) \\ 4.9  \end{array} $ &  
$\begin{array}{@{}c@{\;}c@{}} 5.77 & (3) \\ 3.08 \end{array} $ &  
$\begin{array}{@{}c@{\;}c@{}} 4.30 & (3) \\ 2.35 \end{array} $ &  
$ 0.41^{+0.35}_{-0.27} $ \\                                 
\enddata

\tablecomments{Total measured (i.e. uncorrected for intrinsic reddening)
emission line fluxes, in units of 10$^{-17}$ erg s$^{-1}$ cm$^{-2}$. Random
errors are given below each flux value. Numbers within parenthesis indicate the
type of Gaussian fit performed to the emission lines: (1) Gaussian with 3 free
parameters; (2) two Gaussians with the same width, different amplitude, and
fixed separation; (3) Gaussian with fixed width and position. In the case of
the restricted fits (types 2 and 3), the a priory required fixed values have
been taken from neighbor and similar lines. [O~{\sc iii}] lines for 
GSS073\_1810 are outside the observed spectral range. The remaining void
entries in the table are due to very high sky line residuals. Note that these
numbers have not been corrected for aperture effects.}

\end{deluxetable}


\begin{deluxetable}{lcccc}
\tablecaption{Aperture corrections estimates.\label{table_aperture_corrections}}
\tablewidth{0pt}
\tablehead{
\colhead{Galaxy} & 
\colhead{\makebox[0in][c]{$U_{300}$}} &
\colhead{\makebox[0in][c]{$B_{450}$}} &
\colhead{\makebox[0in][c]{$V_{606}$}} &
\colhead{\makebox[0in][c]{$I_{814}$}} \\
\colhead{                       } & 
\colhead{\makebox[0in][c]{slit 0.75\arcsec}} &
\colhead{\makebox[0in][c]{slit 0.75\arcsec}} &
\colhead{\makebox[0in][c]{slit 0.75\arcsec}} &
\colhead{\makebox[0in][c]{slit 0.75\arcsec}} 
}
\startdata
hd2-264.1     &
$ 0.61 \;(0.52\pm 0.11) $ &   
$ 0.58 \;(0.50\pm 0.12) $ &   
$ 0.58 \;(0.50\pm 0.12) $ &   
$ 0.60 \;(0.51\pm 0.11) $ \\  
hd2-264.2     &
$ 0.96 \;(0.75\pm 0.30) $ &   
$ 0.92 \;(0.72\pm 0.27) $ &   
$ 0.90 \;(0.70\pm 0.26) $ &   
$ 0.88 \;(0.66\pm 0.27) $ \\  
\noalign{\vskip .7ex} \cline{2-5} \noalign{\vskip 1.4ex}
 & 
\makebox[0in][c]{$U_{300}$} &
\makebox[0in][c]{$B_{450}$} &
\makebox[0in][c]{$V_{606}$} &
\makebox[0in][c]{$I_{814}$} \\
 & 
\makebox[0in][c]{slit 1.25\arcsec} &
\makebox[0in][c]{slit 1.25\arcsec} &
\makebox[0in][c]{slit 1.25\arcsec} &
\makebox[0in][c]{slit 1.25\arcsec} \\ \cline{2-5} \noalign{\vskip 0.7ex}
hd4-656.1     &
$ 0.85 \;(0.69\pm 0.17) $ &   
$ 0.84 \;(0.67\pm 0.18) $ &   
$ 0.82 \;(0.67\pm 0.17) $ &   
$ 0.81 \;(0.68\pm 0.15) $ \\  
hd4-795.111   &
$ 0.77 \;(0.63\pm 0.14) $ &   
$ 0.78 \;(0.62\pm 0.17) $ &   
$ 0.81 \;(0.65\pm 0.17) $ &   
$ 0.86 \;(0.68\pm 0.17) $ \\  
\noalign{\vskip .7ex} \cline{2-5} \noalign{\vskip 1.4ex}
 & 
\makebox[0in][c]{$V_{606}$} &
\makebox[0in][c]{$I_{814}$} &
\makebox[0in][c]{$V_{606}$} &
\makebox[0in][c]{$I_{814}$} \\
 & 
\makebox[0in][c]{slit 0.76\arcsec} &
\makebox[0in][c]{slit 0.76\arcsec} &
\makebox[0in][c]{slit 1.18\arcsec} &
\makebox[0in][c]{slit 1.18\arcsec} \\ \cline{2-5} \noalign{\vskip 0.7ex}
GSS073\_1810  &
$ 0.52 \;(0.45\pm 0.10) $ &   
$ 0.56 \;(0.47\pm 0.11) $ &   
$ 0.69 \;(0.65\pm 0.06) $ &   
$ 0.72 \;(0.67\pm 0.07) $ \\  
\noalign{\vskip .7ex} \cline{2-5} \noalign{\vskip 1.4ex}
 & 
\makebox[0in][c]{$V_{606}$} &
\makebox[0in][c]{$I_{814}$} &
\makebox[0in][c]{$V_{606}$} &
\makebox[0in][c]{$I_{814}$} \\
 & 
\makebox[0in][c]{slit 0.76\arcsec} &
\makebox[0in][c]{slit 0.76\arcsec} &
\makebox[0in][c]{slit 1.00\arcsec} &
\makebox[0in][c]{slit 1.00\arcsec} \\ \cline{2-5} \noalign{\vskip 0.7ex}
GSS084\_4515  &
$ 0.88 \;(0.60\pm 0.24) $ &   
$ 0.90 \;(0.67\pm 0.25) $ &   
$ 0.94 \;(0.76\pm 0.22) $ &   
$ 0.97 \;(0.81\pm 0.21) $ \\  
GSS084\_4521  &
$ 0.60 \;(0.50\pm 0.10) $ &   
$ 0.61 \;(0.52\pm 0.11) $ &   
$ 0.73 \;(0.62\pm 0.14) $ &   
$ 0.74 \;(0.65\pm 0.12) $ \\  
\enddata

\tablecomments{Fraction of light inside the slit for all the observational
configurations. These numbers have been measured in the available HST
images (F300W, F450W, F606W and F814W for \mbox{HDF-N} galaxies, and F606W and
F814W for GSS galaxies). The numbers within parenthesis indicate the mean and
standard deviation in numerical simulations, assuming typical errors in the
positioning of the slit on the targets of 0.1\arcsec\ in both right ascension
and declination, and of 3\arcdeg\ in position angle.}

\end{deluxetable}


\begin{deluxetable}{lccccc}
\tablecaption{Additional flux data for 
HDF-N galaxies.\label{table_more_fluxes_hdfn}}
\tablewidth{0pt}
\tablehead{
\colhead{Galaxy} & 
\colhead{$U_{300}$\tablenotemark{a}} &
\colhead{$B_{450}$\tablenotemark{a}} &
\colhead{$V_{606}$\tablenotemark{a}} &
\colhead{$I_{814}$\tablenotemark{a}} &
\colhead{$K$\tablenotemark{a}}
}
\startdata
hd2-264.1     &                     
$24.93 \pm 0.08 $ &                       
$23.52 \pm 0.01 $ &                       
$22.30 \pm 0.01 $ &                       
$21.50 \pm 0.01 $ &                       
$19.93 \pm 0.01 $ \\                      
hd2-264.2     &                     
$24.12 \pm 0.03 $ &                       
$23.48 \pm 0.01 $ &                       
$22.70 \pm 0.01 $ &                       
$22.10 \pm 0.01 $ &                       
$21.43 \pm 0.02 $ \\                      
hd4-656.1     &                     
$22.84 \pm 0.01 $ &                       
$22.22 \pm 0.01 $ &                       
$21.33 \pm 0.01 $ &                       
$20.74 \pm 0.01 $ &                       
$19.91 \pm 0.01 $ \\                      
hd4-795.111   &                     
$23.22 \pm 0.02 $ &                       
$22.55 \pm 0.01 $ &                       
$21.64 \pm 0.01 $ &                       
$21.00 \pm 0.01 $ &                       
$19.92 \pm 0.01 $ \\                      
\enddata

\tablenotetext{a}{Magnitudes are given in the AB system, where
\mbox{$m=-2.5\log f_\nu -48.60$}, with $f_\nu$ in erg cm$^{-2}$
s$^{-1}$ Hz$^{-1}$. All the data extracted from Table~5 of \citet{fer99}.}

\end{deluxetable}


\begin{deluxetable}{lccccccc}
\rotate
\tablecaption{Additional flux data for GSS 
galaxies.\label{table_more_fluxes_gss}}
\tablewidth{0pt}
\tablehead{
\colhead{Galaxy} & 
\colhead{$U_{\rm mag}$\tablenotemark{a}} &
\colhead{$B_{\rm mag}$\tablenotemark{a}} &
\colhead{$R_{\rm mag}$\tablenotemark{a}} &
\colhead{$I_{\rm mag}$\tablenotemark{a,b}} &
\colhead{$V_{606}$\tablenotemark{c}} &
\colhead{$I_{814}$\tablenotemark{c}} &
\colhead{$J_{\rm N3}$\tablenotemark{d}}
}
\startdata
GSS073\_1810\tablenotemark{e} &    
$23.38 \pm 0.02 $ &                       
$23.00 \pm 0.02 $ &                       
$21.76 \pm 0.01 $ &                       
$20.98 \pm 0.01 $ &                       
$22.00 \pm 0.01 $ &                       
$20.91 \pm 0.01 $ &                       
$20.83 \pm 0.28 $ \\                      
GSS084\_4515 &                     
$24.54 \pm 0.04 $ &                       
$24.34 \pm 0.04 $ &                       
$23.24 \pm 0.03 $ &                       
$22.60 \pm 0.04 $ &                       
$23.73 \pm 0.04 $ &                       
$22.78 \pm 0.02 $ &                       
---               \\                      
GSS084\_4521 &                     
$24.08 \pm 0.02 $ &                       
$23.47 \pm 0.02 $ &                       
$21.85 \pm 0.01 $ &                       
$21.04 \pm 0.01 $ &                       
$21.80 \pm 0.02 $ &                       
$20.98 \pm 0.01 $ &                       
---               \\                      
\enddata

\tablenotetext{a}{Magnitudes are given in the AB system, where $m=-2.5\log
f_\nu -48.60$, with $f_\nu$ in erg cm$^{-2}$ s$^{-1}$ Hz$^{-1}$. Data extracted
from \citet{bru99}.}

\tablenotetext{b}{The $I_{\rm mag}$ values have been corrected for the
0.15 mag offset found in the comparison shown in
Fig.~\ref{figure_plot_brunner}.}

\tablenotetext{c}{HST $V606$ and $I814$ magnitudes in the AB system, measured
in this work using SExtractor \citep{ber96}.}

\tablenotetext{d}{NIRSPEC $J$ magnitude (filter N3) measured in this work (see
text for details).}

\tablenotetext{e}{Tabulated magnitudes correspond to the integrated
values of the two interacting galaxies.}

\end{deluxetable}


\begin{deluxetable}{lcc}
\tablecaption{Mid-IR and radio fluxes for the galaxy 
sample.\label{table_more_fluxes_mir_radio}}
\tablewidth{0pt}
\tablehead{
\colhead{Galaxy} & 
\colhead{$S_\nu(15\mu{\rm m})$\tablenotemark{a}} &
\colhead{$S_\nu(\rm radio)$\tablenotemark{b}}
}
\startdata
hd2-264.1     &                     
$  115 \pm 40 $ &                         
$   14 \pm 3 $ \\                         
\noalign{\vskip .5ex}
hd2-264.2     &                     
$<100$ &                                  
$<9$ \\                                   
\noalign{\vskip .5ex}
hd4-656.1     &                     
$ 49^{+36}_{-9} $ &                       
$<9$ \\                                   
\noalign{\vskip .5ex}
hd4-795.111   &                     
$ 52^{+34}_{-9} $ &                       
$<9$ \\                                   
\noalign{\vskip .5ex}
GSS073\_1810\tablenotemark{c} &     
$ 361  \pm 80 $ &                         
$\la 16$ \\                                  
\noalign{\vskip .5ex}
GSS084\_4515  &                     
$<200$ &                                  
$<16$ \\                                  
\noalign{\vskip .5ex}
GSS084\_4521  &                     
$  240 \pm 90 $ &                         
$  24 \pm 4 $ \\                          
\enddata

\tablenotetext{a}{Mid-infrared flux densities, in $\mu$Jy, as measured by
ISOCAM with the LW3 filter. See references in the text.}

\tablenotetext{b}{Radio flux densities in $\mu$Jy, at 8.5~GHz for the HDF-N
galaxies, and at 5.0~GHz for the GSS galaxies. See references in the text.}

\tablenotetext{c}{Tabulated fluxes correspond to the integrated
values of the two interacting galaxies.}

\end{deluxetable}


\begin{deluxetable}{lc@{$\;\;$}cc@{$\;\;$}cc@{$\;\;$}ccc}
\tablecaption{SFR estimates (M$_\odot$/yr) corrected for extinction and
for aperture effects.\label{table_sfrs}}
\tablewidth{0pt}
\tablehead{
\colhead{Galaxy} & 
\colhead{SFR$_{\rm UV}$} &  
A(UV)\tablenotemark{a} &
\colhead{SFR$_{\rm\mbox{\scriptsize [O~{\sc ii}]}}$} & 
A([O~{\sc ii}])\tablenotemark{a} &
\colhead{SFR$_{\rm H\alpha}$} &  
A(${\rm H}\alpha$)\tablenotemark{a} &
\colhead{SFR$_{\rm IR}$} &
\colhead{SFR$_{\rm radio}$}
}
\startdata
hd2-264.1            &          
$  1.2_{-0.5}^{+0.8} $ & $1.5_{-0.5}^{+0.6} $ &     
$  4.4_{-2.0}^{+3.8} $ & $1.7_{-0.6}^{+0.7} $ &     
$  3.6_{-1.1}^{+1.6} $ & $0.88_{-0.32}^{+0.34}$ &     
$   6.4_{-2.4}^{+3.3}$ &              
$    11    \pm   4   $ \\             
\noalign{\vskip .5ex}
hd2-264.2     &                 
$  5.1_{-1.4}^{+2.0} $ & $2.2_{-0.3}^{+0.4} $ &     
$  7.1_{-2.3}^{+3.5} $ & $2.6_{-0.4}^{+0.4} $ &     
$  3.6_{-0.7}^{+0.9} $ & $1.3_{-0.2}^{+0.2} $ &     
$ < 5.2 $              &              
$ < 20 $               \\             
\noalign{\vskip .5ex}
hd4-656.1     &                 
$  2.4_{-0.4}^{+0.5} $ & $0.23_{-0.18}^{+0.19} $ &     
$  2.8_{-0.6}^{+0.7} $ & $0.24_{-0.23}^{+0.23} $ &     
$  1.8_{-0.2}^{+0.2} $ & $0.12_{-0.12}^{+0.12} $ &     
$   2.3_{-0.5}^{+1.7}$ &              
$ < 18 $               \\             
\noalign{\vskip .5ex}
hd4-795.111   &                 
$  2.7_{-0.5}^{+0.7} $ & $0.83_{-0.25}^{+0.25} $ &     
$  1.9_{-0.5}^{+0.6} $ & $0.97_{-0.30}^{+0.31} $ &     
$  2.2_{-0.4}^{+0.4} $ & $0.49_{-0.15}^{+0.15} $ &     
$   2.1_{-0.2}^{+1.4}$ &              
$ < 16 $               \\             
\noalign{\vskip .5ex}
GSS073\_1810A&                  
 ---                   &        &     
$ 35_{-13}^{+22}$ &      $2.1_{-0.5}^{+0.5} $ &     
$ 29.8_{-6.9}^{+9.8} $ & $1.1_{-0.2}^{+0.3} $ &     
 ---                   &              
 ---                   \\             
\noalign{\vskip .5ex}
GSS073\_1810B&                  
 ---                   &        &     
$  7.4_{-3.8}^{+8.3} $ & $1.1_{-0.7}^{+0.8} $ &     
$  9.4_{-3.1}^{+5.2} $ & $0.58_{-0.37}^{+0.42} $ &     
 ---                   &              
 ---                   \\             
\noalign{\vskip .5ex}
GSS073\_1810\tablenotemark{b} & 
$ 15.0_{-3.8}^{+5.2} $ & $1.4_{-0.3}^{+0.3} $ &     
$ 42_{-13}^{+23}$ &      $1.7_{-0.4}^{+0.3} $ &     
$ 39_{-8}^{+11} $ &      $0.84_{-0.2}^{+0.2} $ &     
$ 161_{-55}^{+50}    $ &              
$ \la 86 $               \\             
\noalign{\vskip .5ex}
GSS084\_4515 &                  
$  3.1_{-1.4}^{+2.6} $ & $0.89_{-0.66}^{+0.67} $ &     
$ 20_{-10}^{+22}$ &      $1.0_{-0.8}^{+0.8} $ &     
$  7.4_{-2.7}^{+4.3} $ & $0.53_{-0.4}^{+0.4} $ &     
$ < 67 $               &              
$ < 81 $               \\             
\noalign{\vskip .5ex}
GSS084\_4521 &                  
$  7.6_{-4.7}^{+17.9}$ & $1.6_{-1.0}^{+1.4} $ &     
$ 15_{-10}^{+50}$ &      $1.9_{-1.2}^{+1.7} $ &     
$  15_{-7}^{+17}$ &      $0.96_{-0.62}^{+0.83} $ &  
$    67_{-35}^{+29}   $ &             
$    94    \pm 44    $ \\             
\enddata

\tablenotetext{a}{Extinction correction factors (in magnitudes) derived from
the color excesses given in Tables~\ref{table_sample} and~\ref{table_fluxes},
and applied to correct the UV, [O~{\sc ii}] and H$\alpha$ fluxes listed in
Table~\ref{table_fluxes} in order to get the SFR values quoted in this table.}

\tablenotetext{b}{SFR$_{\rm\mbox{\scriptsize [O~{\sc ii}]}}$ and
SFR$_{\rm H\alpha}$ values for this object correspond to the sum of the 
respective values for the individual galaxies (listed in the previous two
entries).}
 
\tablecomments{See Fig.~\ref{figure_sfrs} for a graphical comparison of these
numbers.}

\end{deluxetable}


\begin{deluxetable}{lccccccc}
\tabletypesize{\footnotesize}
\rotate
\tablecaption{Emission line ratios and 
metallicities.\label{table_emission_line_ratios}}
\tablewidth{0pt}
\tablehead{
\colhead{Galaxy} &
\colhead{$\frac{\mbox{[O~{\sc iii}]$\lambda$5007}}{\mbox{H$\beta$}}$} &
\colhead{$\frac{\mbox{[N~{\sc ii}]$\lambda$6583}}{\mbox{H$\alpha$}}$} &
\colhead{$\frac{\mbox{[S~{\sc ii}]$\lambda\lambda$6716,6731}}%
{\mbox{H$\alpha$}}$} &
\colhead{$\frac{\mbox{[N~{\sc ii}]$\lambda$6583}}%
{\mbox{[O~{\sc ii}]$\lambda$3727}}$} &
\colhead{R$_{23}$\tablenotemark{a}} &
\colhead{O$_{32}$\tablenotemark{b}} &
\colhead{$12+\log(\mbox{O/H})$\tablenotemark{c}}
}
\startdata
hd2-264.1     & 
$1.14\pm0.16$ & $0.22\pm0.06$ & $0.13\pm0.07$ & $0.32\pm0.10$ &
$3.40\pm0.55$ & $0.75\pm0.14$ & $8.83\pm0.06$ \\
hd2-264.2     & 
$> 0.21$      & $0.24\pm0.05$ & $0.13\pm0.05$ & $0.22\pm0.04$ &
$> 3.47$      & $> 0.09$      & $8.78\pm0.06$ \\
hd4-656.1     & 
$0.41\pm0.21$ & $0.27\pm0.07$ & $0.32\pm0.06$ & $0.31\pm0.08$ &
$3.09\pm0.29$ & $0.22\pm0.09$ & $8.84\pm0.04$ \\
hd4-795.111   & 
$0.21\pm0.06$ & $0.32\pm0.07$ & $0.17\pm0.03$ & $0.67\pm0.15$ &
$1.66\pm0.16$ & $0.20\pm0.05$ & $9.00\pm0.02$ \\
GSS073\_1810A &
---           & $0.37\pm0.06$ & $0.32\pm0.04$ & $0.43\pm0.07$ &
---           & ---           & ---           \\
GSS073\_1810B &
---           & $0.44\pm0.09$ & $0.33\pm0.09$ & $0.75\pm0.14$ &
---           & ---           & ---           \\
GSS084\_4515  & 
$3.41\pm0.50$ & $0.11\pm0.08$ & $0.13\pm0.12$ & $0.07\pm0.05$ &
$9.48\pm1.37$ & $1.03\pm0.05$ & $8.33\pm0.11$ \\
GSS084\_4521\tablenotemark{d}&
$0.29\pm0.23$ & $0.57\pm0.28$ & $0.55\pm0.22$ & $0.83\pm0.40$ &
$2.34\pm0.81$ & $0.19\pm0.14$ & $8.92\pm0.10$
\enddata

\tablenotetext{a}{R$_{23} \;\equiv\;$
([O~{\sc ii}]$\lambda$3727+[O~{\sc iii}]$\lambda\lambda$4959,5007)/H$\beta$}

\tablenotetext{b}{O$_{32} \;\equiv\;$
([O~{\sc iii}]$\lambda\lambda$4959,5007)/[O~{\sc ii}]$\lambda$3727}

\tablenotetext{c}{Errors correspond to the propagation of the emission line
fluxes uncertainties when deriving the metallicities from the R$_{23}$ and
O$_{32}$ indices (see Fig.~\ref{figure_metallicities}a). 
However, additional uncertainties
associated to the use of this empirical calibration are $\pm 0.15$~dex
\citep{kob99}.}

\tablenotetext{d}{The [O~{\sc iii}]$\lambda$5007 flux for this galaxy has been
determined from [O~{\sc iii}]$\lambda$4959.}

\tablecomments{All the emission line fluxes have been corrected for
extinction.}

\end{deluxetable}


\begin{deluxetable}{lcccc}
\tablecaption{Additional galaxy properties.\label{table_more_galaxy_properties}}
\tablewidth{0pt}
\tablehead{
\colhead{Galaxy} & 
\colhead{$M_B$\tablenotemark{a}} &
\colhead{$R_{\rm e}$ (kpc)\tablenotemark{b}} &
\colhead{Mass (M$_\odot$)\tablenotemark{c}} &
\colhead{$L_{\rm IR}/L_\odot$\tablenotemark{d}}
}
\startdata
hd2-264.1     & 
$-$19.7 &
$5.69_{-0.24}^{+2.63}$ &
$4.8\times 10^{10}$ &
$10^{10.57}$ \\
hd2-264.2     & 
$-$19.4 &
$1.02_{-0.40}^{+0.44}$ &
$1.2\times 10^{10}$ &
\makebox[0in][r]{$<\,$}$10^{10.48}$ \\
hd4-656.1     & 
$-$20.6 &
$3.29_{-0.11}^{+0.49}$ &
$4.4\times 10^{10}$ &
$10^{10.13}$ \\
hd4-795.111   & 
$-$20.1 &
$2.63_{-0.21}^{+0.93}$ &
$3.9\times 10^{10}$ &
$10^{10.09}$ \\
GSS073\_1810A & 
$-$21.3 &
$5.76_{-0.08}^{+0.11}$ &
--- &
--- \\
GSS073\_1810B & 
$-$21.4 &
$3.92_{-0.10}^{+0.27}$ &
--- &
--- \\
GSS073\_1810  & 
$-$22.1 &
---  &
--- &
$10^{11.97}$ \\
GSS084\_4515  & 
$-$20.1 &
$2.45_{-0.12}^{+0.28}$ &
--- &
\makebox[0in][r]{$<\,$}$10^{11.59}$ \\
GSS084\_4521  & 
$-$21.7 &
$4.24_{-0.05}^{+0.27}$ &
--- &
$10^{11.59}$
\enddata

\tablenotetext{a}{Rest-frame absolute magnitudes in the $B$~band (AB system),
derived as explained in Sect.~\ref{galaxy_properties}.}

\tablenotetext{b}{Half-light radii (in kpc), computed as explained in
Sect.~\ref{galaxy_properties}.}

\tablenotetext{c}{Galaxy masses (in solar units) derived as explained in
Sect.~\ref{galaxy_properties}}

\tablenotetext{d}{Total IR luminosity, \mbox{$L_{\rm IR} = L(\mbox{8--1000}\mu
m)$}, in solar units. These numbers have been derived as explained in
Sect.~\ref{subsection_sfr_ir}.}

\end{deluxetable}


\clearpage


\clearpage


\clearpage


\clearpage


\clearpage


\end{document}